\documentclass[aps,twocolumn,superscriptaddress]{revtex4}
\usepackage{graphicx,subfigure}
\usepackage{subfig}
\usepackage{amsthm}
\usepackage{amsmath}
\usepackage{color,colordvi}

\begin{document}
\bibliographystyle{revtex4}


\title{The role of surface  quenching of the singlet delta molecule in a capacitively coupled oxygen discharge
}



\author{A. Proto}

\affiliation{Science Institute, University of Iceland,
                Dunhaga 3, IS-107 Reykjavik, Iceland}

\author{J. T. Gudmundsson}
\email[]{tumi@hi.is}


\affiliation{Science Institute, University of Iceland,
                Dunhaga 3, IS-107 Reykjavik, Iceland}

\affiliation{Department of Space and Plasma Physics, School of Electrical Engineering and Computer Science, 
KTH--Royal Institute of Technology, SE-100 44, Stockholm, Sweden}











\date{\today}

\begin{abstract}
We use the one-dimensional object-oriented particle-in-cell Monte Carlo collision code {\tt oopd1}
to explore the influence of the surface quenching of the singlet delta metastable molecule O$_2$(a$^1\Delta_{\rm g}$) 
on the 
 electron heating mechanism, and the electron energy probability function
(EEPF), in  a single frequency   capacitively coupled oxygen discharge.   
When operating at low pressure (10 mTorr) varying  the surface quenching coefficient
in the range 0.00001 -- 0.1 has no influence on the 
electron heating mechanism and electron heating is dominated by drift-ambipolar (DA) heating in the plasma bulk
and electron cooling is observed in the sheath regions. As the pressure is increased to 25 mTorr the electron heating
becomes a combination of DA-mode and $\alpha-$mode heating, and the role of the DA-mode decreases with decreasing 
surface quenching coefficient.
At 50 mTorr electron heating in the sheath region dominates.  However, for the highest quenching 
coefficient there is 
some contribution from the DA-mode in the plasma bulk, but this contribution decreases to almost 
zero and  pure 
$\alpha-$mode electron heating is observed for a surface quenching coefficient of 0.001 or smaller.

\end{abstract}
\pacs{52.50.Pi,52.57.-j,52.50.Nr,52.65.Rr,82.33.Xj}

\maketitle


\section{Introduction}

Low pressure radio frequency (rf) driven capacitively coupled plasma (CCP) 
discharges have been applied in integrated circuit manufacturing for a few decades. 
Currently the  CCPs consist of two parallel electrodes, typically of radius of a few tens 
of cm, separated by a few cm,  and driven by a radio-frequency power supply. 
In the capacitively coupled discharges a plasma forms between the electrodes, from which it is separated by space charge sheaths. 
The energy transport mechanism and particle interactions in the plasma-surface interface region play
a significant role in the discharge operation. Atomic species recombine to form molecules
  and metastable species are
quenched on the electrode surfaces.  Both of these processes influence the discharge operation and can
have determining influence on the electronegativity of the discharge and the electron heating mechanisms and
thus on the electron kinetics.   In turn the electron kinetics dictate the ionization and dissociation processes
that maintain the discharge and create the radicals that are desired for materials processing.

When operated at low pressure the electron heating mechanism in a CCP is referred to as being
collisionless and is associated with the electron dynamics in the sheath region,
a rapid movement of the electrode sheaths or  stochastic  electron heating
\citep{lieberman98:955,gozadinos01:117}.  When the  electrons interact  with the moving sheaths, they can
be either cooled (collapsing sheath) or heated (expanding sheath).
Energetic electrons can also bounce back and forth between the two sheaths.  When they 
hit the sheath edge during its expansion phase, energy is transferred to the electrons.  
 This electron heating process is referred to as  electron bounce resonance heating (BRH) 
and can occur for certain combinations of driving frequency and electrode gap 
\citep{wood91t,wood95:89,liu11:055002,liu12:114101,wilczek15:024002}.
The sheath motion and thus the stochastic heating can also be enhanced by self-excited non-linear plasma
series resonance (PSR) oscillations  
\citep{czarnetzki06:123503,donko09:131501,schungel15:044009,wilczek16:063514}.
Collisionless electron heating via sheath oscillations  is commonly referred to as the $\alpha$-mode
\citep{belenguer90:4447}.
When the discharge is operated at  high applied voltages and pressures secondary 
electron emission can contribute to or even dominate the ionization, and the  operating mode is
then referred to as $\gamma$-mode \citep{belenguer90:4447}.
In electronegative discharges large electron density gradients can develop within the rf period
which can cause the generation of ambipolar fields along with drift fields, that can accelerate the electrons, a heating mechanism  referred to as the drift-ambipolar (DA) mode 
\citep{schulze11:275001,derzsi15:346}.

The oxygen discharge has been applied in plasma materials 
processing for decades and its applications  include processes such as
oxidation or anodization of silicon \citep{pulfrey73:1529,kawai94:2223,hess99:127}, 
ashing of photoresist \citep{tolliver84:1,hartney89:1}, 
and surface modification  of polymer films \citep{vesel12:634,chashmejahanbin14:44,vesel17:293001}.
The oxygen discharge is  weakly electronegative and the electronegativity 
depends on the control parameters including pressure and power \citep{gudmundsson01:1100}.  
At low operating pressure the negative O$^-$-ion is the dominant negative ion and it
is created almost solely by electron impact dissociative attachment,
where the singlet metastables play a significant role \citep{gudmundsson07:399,toneli15:325202}.
 Earlier we have demonstrated how these singlet metastable molecular states influence the 
 electron heating mechanism and thus the electron kinetics in the capacitively coupled oxygen 
discharge operated at a single frequency of 13.56 MHz 
\citep{gudmundsson15:035016,gudmundsson15:153302,hannesdottir16:055002,gudmundsson17:120001} 
as well as the ion energy  distribution in both single and dual frequency discharges
\citep{hannesdottir17:175201}.   We have demonstrated that at low pressure (10 mTorr), 
the  electron heating is mainly within the  plasma bulk 
(the electronegative core), and  at higher pressures (50 –-- 500 mTorr) 
the electron heating occurs mainly in the  sheath region
 \citep{gudmundsson15:153302,hannesdottir16:055002}. 
When operating at low pressure  the electron 
heating within the discharge is due to 
combined drift-ambipolar-mode (DA-mode) and $\alpha$-mode and at higher pressures the discharge is
operated in the  $\alpha$-mode \citep{gudmundsson17:193302,gudmundsson18:025009}.

Recent fluid model and PIC/MCC simulation studies have
 indicated that there are significant changes in the 
 electronegativity and the electron heating mechanism 
 as the quenching coefficient for the O$_2$(a$^1\Delta_{\rm g}$) on the electrode
surfaces is  varied \citep{greb15:044003,derzsi16:015004,derzsi17:034002}. 
\citet{derzsi16:015004} using a PIC/MCC simulation  demonstrate that the 
O$_2$(a$^1\Delta_{\rm g}$) density decreases exponentially with increasing quenching coefficient 
$\gamma_{\rm wqa}$ in the range $10^{-4} \leq \gamma_{\rm wqa} \leq 5 \times 10^{-2}$.
 In these PIC/MCC simulation  studies  \citep{derzsi16:015004,derzsi17:034002}  
the    O$_2$(a$^1\Delta_{\rm g}$) density is taken as a fraction of the 
ground state oxygen molecule  O$_2$(X$^3\Sigma_{\rm g}^-)$.
Similarly, using a 1D fluid model,   \citet{greb15:044003} demonstrated  
that the electronegativity depends strongly 
on the  O$_2$(a$^1\Delta_{\rm g}$)  surface quenching coefficient and  argued
 that  increased quenching coefficient leads to 
decreased  O$_2$(a$^1\Delta_{\rm g}$) density,  decreased detachment by the   O$_2$(a$^1\Delta_{\rm g}$) 
state, and thus higher negative ion density.  This is due to the very effective annihilation of
the O$^-$-ions in the plasma bulk via detachment by the singlet metastable molecules O$_2$(a$^1\Delta_{\rm g}$).
More recently \citet{gibson17:115007} explored the particle dynamics in an oxygen CCP 
while keeping the  O$_2$(a$^1\Delta_{\rm g}$) as either 16 \% or 0.5 \% of the ground state density, 
to create a weekly and highly, electronegative oxygen discharge, respectively. Using a 1D fluid model they  demonstrated
that oxygen discharges can operate in distinctly different modes dependent upon the  O$_2$(a$^1\Delta_{\rm g}$)  density 
within the discharge. 
Less is known about the role of the quenching of  O$_2$(b$^1\Sigma_{\rm g}$) on the electrodes.

Here we study  how the surface  quenching coefficients for the singlet metastable molecules
O$_2$(a$^1\Delta_{\rm g}$)
 influence 
the electron heating processes, the electron energy probability function (EEPF), 
the effective electron temperature, 
in the single frequency voltage driven capacitively coupled oxygen discharge by means of numerical
simulation, for a fixed discharge voltage, while the discharge pressure is varied from 10 to 50  mTorr.    
The simulation parameters and the cases explored are defined in section \ref{cases}, where we give an 
overview of the known surface quenching coefficients for the singlet metastable molecules on various 
surfaces and determine the partial pressures of the neutral background species using a global model.
 The results of the PIC/MCC simulations, the various electron heating modes observed for the  various combinations
of surface quenching coefficients and operating pressures,  are  discussed  
in section \ref{results}.  
Finally, concluding remarks are given in section \ref{conclusion}.


\section{The simulation}
\label{cases}

The one-dimensional object-oriented particle-in-cell Monte Carlo collision (PIC/MCC) code {\tt oopd1}
\citep{hammel03:66, verboncoeur95:199} is here applied to a capacitively coupled oxygen discharge.
In 1d-3v PIC codes, like  {\tt oopd1}, the model system has one spatial dimension and three velocity components.
In our earlier work we added oxygen atoms in the ground state O$(^3$P) and ions of the oxygen 
atom O$^+$ and the relevant reactions to the {\tt oopd1} discharge model \citep{gudmundsson13:035011}. 
Later we added the singlet  metastable molecule O$_2$(a$^1\Delta_{\rm g}$),
the metastable oxygen atom  O$(^1$D)  \citep{gudmundsson15:035016}, and  the singlet
metastable molecule O$_2$(b$^1\Sigma_{\rm g}^+$) \citep{hannesdottir16:055002}, along with
 energy dependent secondary electron emission coefficients
for oxygen ions and neutrals  as they bombard the electrodes \citep{hannesdottir16:055002}.
For this current work the discharge model contains nine species: electrons, the ground state 
neutrals O($^3$P) 
and O$_2(\mathrm{X}^3\Sigma_{\rm g}^-)$, the negative ions O$^-$, the positive ions O$^+$ and O$_2^+$, and the metastables 
O($^1$D), O$_2(\mathrm{a}^1\Delta_{\rm g})$ and O$_2$(b$^1\Sigma_{\rm g}^+)$.
The full oxygen reaction set and the cross sections used have been discussed in our earlier works 
and will not be repeated here \citep{hannesdottir16:055002, gudmundsson15:035016, gudmundsson13:035011}. 
 However, as the role of the singlet  metastable oxygen molecule
O$_2$(a$^1\Delta_g$) is being explored two important reactions are mentioned here (see further discussion in \citet{gudmundsson15:035016}).
  It is known from global model studies \cite{gudmundsson07:399} that dissociative attachment 
of the oxygen molecule is almost the sole source of O$^-$-ions in the discharge and 
 the  metastable oxygen molecules play a major role.  In particular 
dissociative attachment from the metastable oxygen molecule
O$_2$(a$^1\Delta_g$) can be the dominant path for the creation of negative ion O$^-$ through
 \[
e + \mathrm{O}_2(\mathrm{a}^1\Delta_g) 
\longrightarrow   \left\{ \begin{array}{l}
                                      \mbox{O}(^3\mbox{P}) +   \mathrm{O}^-   \\
                                         \mbox{O}(^1\mbox{D}) +  \mathrm{O}^-
\end{array} \right.
\]
Lower pressure and thus higher effective electron temperature promotes the creation of the negative ion O$^-$.
The metastable  molecule O$_2$($\mathrm{a}^1\Delta_g$) also contributes significantly to the loss
of the negative ion O$^-$-ion through the detachment process
\[
\mathrm{O}^- +  \mathrm{O}_2(\mathrm{a}^1\Delta_g) \longrightarrow
 \mathrm{products}
\]
while detachment by the oxygen molecule in the ground state is negligible.
  Here we use the rate coefficient
measured at 400 K of $1.5 \times 10^{-16}$ m$^{3}$/s by  \citet{midey08:3040} to estimate the cross
section by assuming a Maxwellian velocity distribution of the particles. The
cross section is allowed to fall as $1/\sqrt{{\cal E}}$ to 184 meV and then
take a fixed value of $5.75 \times 10^{-20}$ m$^2$.  Also we assume that the  detachment by
 the metastable molecule  O$_2$($\mathrm{a}^1\Delta_g$) leads to the formation of O($^3$P) +
 O$_2(\mathrm{X}^3 \Sigma_g^-)$ + e, instead of O$_3$ + e and O + O$_2^-$. Increased discharge pressure thus 
promotes the loss of the negative ion  O$^-$.  

We assume a symmetric  capacitively coupled discharge where one of the electrodes is driven by an rf voltage
\begin{equation}
V(t) = V_0 \sin( 2 \pi f t)
\end{equation}
while the other is grounded.  Here $V_0$ is the voltage amplitude, $f$ the driving frequency,
 and $t$ is the time. 
For this current study we assume the discharge to be 
operated with voltage amplitude of  $V_0 = 222$ V with an electrode
 separation of 4.5 cm and a capacitor of 1 F in series with the voltage source, while the surface quenching 
coefficient for the singlet delta metastable  O$_2(\mathrm{a}^1\Delta_{\rm g})$ and discharge pressure is varied. 
 These are the same parameters
 as assumed in our earlier work using {\tt oopd1} 
\citep{gudmundsson13:035011,gudmundsson15:035016,hannesdottir16:055002,gudmundsson17:120001,gudmundsson18:025009} and
by  Lichtenberg {\it et al.}~\citep{lichtenberg94:2339} using the {\tt xpdp1} code.
 The discharge electrode separation 
is assumed to be small compared to the electrode diameter so that the discharge can be
treated as one dimensional. 
We assume 10.25 cm diameter electrodes in order to determine the absorbed power and set the discharge volume
 for the global model calculations, discussed in section \ref{global}.  
The time step $\Delta t$ and the grid spacing $\Delta x$  resolve the electron plasma frequency and the
electron Debye length of the low-energy electrons, respectively, according to
$\omega_{\rm pe} \Delta t <  0.2$, where $\omega_{\rm pe}$ is the electron plasma frequency, and the simulation
grid is uniform and consists of 1000 cells. The electron time step is $3.68 \times 10^{-11}$ s.  
The simulation was run for $5.5 \times 10^6$ time steps or 2750 rf cycles.   It takes roughly  
1700 rf cycles to reach equilibrium for all particles and  
the time averaged plasma parameters shown, such as the  densities, the electron heating rate, 
and the effective electron temperature, are averages over 1000 rf cycles. 
All particle interactions are treated by the Monte Carlo method with a null-collision scheme \citep{birdsall91:65}.
For the heavy particles we use a sub-cycling and the heavy particles are advanced every  16 electron time steps
 and we assume that the initial density profiles are parabolic  \citep{kawamura00:413}.

The kinetics of the charged  particles (electrons, O$^+_2$-ions, O$^+$-ions  and O$^-$-ions) was followed for all energies.
Since the neutral gas density is much higher than the densities of charged species, 
the neutral species at thermal energies (below a certain cut-off energy) 
are treated as a background with fixed
density and temperature and maintained uniformly in space. 
These neutral background species are assumed to have a
Maxwellian velocity distribution at the gas temperature (here~$\mathrm{T}_{\mathrm{n}}$ = 26 mV).  
The kinetics of the neutrals are followed when their energy exceeds a preset energy threshold value.
The energy threshold values used here for the various neutral species are listed in Table \ref{simpar}. 
 Due to recombination of atomic oxygen and quenching of metastable atoms and molecules on the electrode surfaces
there is a drop in the high energy (energy above the threshold value)  atomic oxygen density and increase in the high energy 
 oxygen molecule densities next to the electrodes
as shown in our earlier work \citep{hannesdottir16:055002}. Thus assuming uniformity of the background 
gas is thus somewhat unrealistic assumption. 

The ratio of the number of physical
particles to computational particles, the particle weight, is also listed in  Table \ref{simpar} 
for all the  neutral species.  Note that in
{\tt  oopd1} the particles can have different  weights  \citep{nguyen06, lim07} and  the  collisions  among  particles  with
different weights is implemented in {\tt oopd1} following the  method  suggested  by  \citet{miller94:1735}.

In our earlier studies we have used fixed partial pressure for each of the neutral species as we
have varied the pressure \citep{gudmundsson15:153302}, the driving voltage amplitude  \citep{gudmundsson17:193302}, 
and the driving frequency \citep{gudmundsson18:025009}. Here we take a different approach and 
calculate the partial pressure for each combination of pressure and surface quenching coefficient using a global
(volume averaged) model  as discussed in section \ref{global}.  
The two electrodes are assumed to be  identical, 
  and the surface coefficients, surface recombination and surface quenching, are kept the same  at both 
electrodes.  We neglect the reflection of electrons from the electrodes.
\begin{table}
\caption{The parameters of the simulation, the particle weight, and the  energy threshold above 
which dynamics of the neutral particles are followed. 
}
 \label{simpar}
 \begin{tabular}{lll}
\hline \hline
Species & particle  & energy    \\
        &  weight   &   threshold    \\
        &                 &  [meV]         \\
\hline
 O$_2(\mathrm{X}^3 \Sigma_{\rm g}^-)$     &  $5 \times 10^7$  & 500                \\    
 O$_2$($\mathrm{a}^1\Delta_{\rm g}$)      &  $5 \times 10^6$  & 100     \\    
 O$_2$($\mathrm{b}^1\Sigma_{\rm g}$)      &  $5 \times 10^6$  & 100     \\    
  O($^3$P)                          &  $5 \times 10^7$  & 500         \\    
  O($^1$D)                          &  $5 \times 10^7$  & 50   \\ 
      O$_2^+$                       &   $10^7$          & -        \\ 
      O$^+$                         &   $10^6$          & -       \\ 
      O$^-$                         &   $5 \times 10^7$ & -      \\ 
      e                            &  $1 \times 10^7$  & -      \\ 
\hline \hline         
 \end{tabular} 
\end{table}

\subsection{Wall recombination coefficients}

As a neutral species hits the electrode it returns as a thermal particle with a given probability
and atoms can recombine to form a thermal molecule with the given probability.
The wall recombination coefficient  for the neutral
 atoms in ground state O($^3$P) is taken to be 0.5 as measured by 
\citet{booth91:611} for a pure oxygen discharge in a stainless steel
reactor at 2 mTorr.  As the oxygen atom O($^3$P) hits the electrode we assume that half of the atoms 
are reflected  as O($^3$P) at room temperature
 and the other half recombines to form the ground state oxygen molecule  O$_2(\mathrm{X}^3 \Sigma_{\rm g}^-)$ at room temperature.
 Note that this is a rough assumption as it
is known that the wall recombination coefficient drops significantly with 
increased pressure \citep{gudmundsson07:399}. 
This could lead to underestimation of the atomic oxygen density. However, the atomic oxygen
density is low and is expected to decrease with increased pressure so this is not expected to  
have a  significant influence on the results reported  here. 
Similarly, as the metastable atom O($^1$D) hits the electrode we assume that half of the atoms are
quenched to form O($^3$P) and the other half recombines to form the ground state oxygen molecule  O$_2(\mathrm{X}^3 \Sigma_{\rm g}^-)$ at room temperature.

\subsection{Wall quenching coefficients}

It is difficult to determine an actual value for the  surface quenching coefficients 
of the singlet metastables on the electrode surfaces 
either  experimentally or theoretically.  In general we would expect that the quenching probability for any
excited species hitting the electrodes to  depend not only on the species itself, 
but also on the surface material, the surface temperature, and the actual
surface condition, such as surface roughness and contamination, which can vary substantially.
Indeed it has been pointed out by \citet{du11:256} that the quenching probability of
 O$_2$(a$^1\Delta_{\rm g}$) increases 
with both the duration of the exposure to  and the concentration of  O$_2$(a$^1\Delta_{\rm g}$).
The values for the measured  wall quenching 
coefficient, found in the literature, for O$_2$(a$^1\Delta_{\rm g}$) and O$_2$(b$^1\Sigma_{\rm g}^+$) on various surfaces 
 are listed in Table \ref{quenchingcoeff}.    All of the values listed in Table \ref{quenchingcoeff}
  were measured at room temperature. We note
that the listed values  span a few orders of magnitude and depend on the surface material. Furthermore, we note
that the measured values also vary
by orders of magnitude for  the same materials.  
We also note that the quenching probability for O$_2$(b$^1\Sigma_{\rm g}^+$) is in general significantly higher than for
 O$_2$(a$^1\Delta_{\rm g}$). 
In our earlier studies 
\citep{toneli15:325202,gudmundsson15:153302,hannesdottir16:055002,gudmundsson17:120001,gudmundsson18:025009} 
we have  used a quenching coefficients  for the singlet metastable 
O$_2$(a$^1\Delta_{\rm g}$) on the electrode surface of $\gamma_{\rm wqa} = 0.007$, estimated by Sharpless and Slanger 
 for iron 
\citep{sharpless89:7947}.  As the measured  wall quenching probability for  O$_2$(a$^1\Delta_{\rm g}$) on aluminum is lower than for iron,
as seen in Table \ref{quenchingcoeff}, 
we would expect that aluminum electrodes would therefore lead to higher 
singlet metastable densities and lower electronegativity.
 In these studies we assumed the quenching
 coefficient for O$_2$(b$^1\Sigma_{\rm g}^+$) to be $\gamma_{\rm wqb} = 0.1$, an assumed value,  based on the suggestion that the 
quenching coefficient  for the 
b$^1\Sigma_{\rm g}^+$ state is about two orders of magnitude larger than for the a$^1\Delta_{\rm g}$ state \citep{obrien70:3832}.
We will use this value for the surface quenching coefficient of  O$_2$(b$^1\Sigma_{\rm g}^+$) in this current study.  We are aware that this may 
be overestimation based on the values listed in Table \ref{quenchingcoeff}.
We have seen in global model studies that wall quenching can be the main loss mechanism for the singlet metastable
state   O$_2$(b$^1\Sigma_{\rm g}^+$) \citep{toneli15:325202}. 
\citet{gordiets95:750} and   \citet{kutasi10:175201} use  wall quenching 
coefficient for O$_2$(a$^1\Delta_{\rm g}$) of  $2 \times 10^{-5}$ and  for O$_2$(b$^1\Sigma_{\rm g}^+$) of  $2 \times 10^{-2}$ for a quartz tube
in their models of flowing N$_2$/O$_2$ dc glow discharge and  Ar/O$_2$ surface-wave microwave discharge, respectively.
By comparing the 1D fluid simulations to phase and space resolved optical emission (PROES)
measurements \citet{greb13:244101} determine the wall quenching coefficient for O$_2$(a$^1\Delta_{\rm g}$) to be 
$1 \times 10^{-5}$ for stainless steel and $3 \times 10^{-3}$ for teflon. A comparison of PIC/MCC simulation with 
experimental findings using PROES for a CCP with aluminum electrodes and electrode spacing of
 $2.5$ cm suggests a wall quenching coefficient of 0.006 \citep{derzsi17:034002}. 
\begin{table}
\caption{Overiew of  the measured  wall quenching coefficients for the 
singlet metastables O$_2$($\mathrm{a}^1\Delta_{\rm g}$) and  O$_2$($\mathrm{b}^1\Sigma_{\rm g}$)
 that can be found in the literature. All of the values listed were measured at room temperature. }
 \label{quenchingcoeff}
 \begin{tabular}{lllll}
\hline \hline
Surface \ \   & O$_2$($\mathrm{a}^1\Delta_{\rm g}$)   &   Ref.  &  O$_2$($\mathrm{b}^1\Sigma_{\rm g}$)  &  Ref.  \\
              & $\gamma_{\rm wqa}$   &    &        $\gamma_{\rm wqb}$ & \\       
\hline
Pyrex & $(3.1 \pm 0.2) \times 10^{-5}$   &\citep{crannage93:267}  & $2 \times 10^{-3}$ & \citep{perram92:427} \\
 & $1.3 \times 10^{-5}$   & \citep{steer69:843}  & $2.2 \times 10^{-3}$ & \citep{arnold67:2053} \\
 & $2.1 \times 10^{-5}$   & \citep{clark69:93} & $1 \times 10^{-2}$ & \citep{izod68:81} \\
 & $4.3 \times 10^{-5}$    & \citep{leiss78:211}     &   & \\
Teflon &  $<  10^{-3}$ &\citep{sharpless89:7947}   &$4.5 \times 10^{-3}$  &\citep{perram92:427}  \\
Fe  & $7 \times 10^{-3}$  & \citep{sharpless89:7947} &  &  \\
  & $4.4 \times 10^{-3}$  & \citep{ryskin81:41} &  &  \\
Cu & $1.4 \times 10^{-2}$ & \citep{sharpless89:7947} & $1.0 \times 10^{-2}$ & \citep{perram92:427}\\
   & $8.5 \times 10^{-4}$ &\citep{ryskin81:41}  & &  \\
 &$2.9 \times 10^{-4}$  &\citep{crannage93:267} & & \\
&$1.15 - 1.43  \times 10^{-3}$  &\citep{du11:256} & & \\
Ni  & $2.7 \times 10^{-3}$ & \citep{ryskin81:41} & $2.6 \times   10^{-2}$ & \citep{perram92:427}  \\
     & $1.1 \times   10^{-2}$ &\citep{sharpless89:7947} & & \\
       &$(3.1 \pm 0.2) \times 10^{-4}$  &\citep{crannage93:267} & & \\
&$0.6 - 1.0  \times 10^{-3}$  &\citep{du11:256} & & \\
Monel (Cu/Ni) & $(2.8 \pm 0.2) \times 10^{-4}$  &\citep{crannage93:267} & & \\
              &  $1.2 \times   10^{-2}$ &\citep{sharpless89:7947} & & \\
Al  & $<  10^{-3}$ &\citep{sharpless89:7947}  &   &  \\
  & $5.9 \times 10^{-5}$ & \citep{ryskin81:41} & &     \\
Pt  & $1 \times   10^{-2}$ &\citep{sharpless89:7947}  &   &  \\
  &  $4.0 \times 10^{-4}$  & \citep{ryskin81:41}  &   &  \\
Ti  & $6.5 \times 10^{-5}$ & \citep{ryskin81:41} & &    \\
    & $<  10^{-3}$ &\citep{sharpless89:7947}  &   &  \\
Ag  &  $1.1 \times 10^{-2}$  & \citep{ryskin81:41}  &   &  \\
Si  &  $7.3 \times 10^{-4}$  & \citep{ryskin81:41}  &   &  \\
Graphite & $3 \times 10^{-3}$  &\citep{sharpless89:7947} & & \\
\hline \hline         
 \end{tabular} 
\end{table}
 Due to the fact that the measured surface quenching coefficients for 
O$_2$($\mathrm{a}^1\Delta_{\rm g}$) vary a few orders of magnitude from roughly $10^{-5}$ to a few times  $10^{-2}$ we allow the
 surface quenching coefficients for 
O$_2$($\mathrm{a}^1\Delta_{\rm g}$) to vary in the range 0.00001 -- 0.1 as we explore the how it influences the 
electron heating processes.   The lowest value of 0.00001 corresponds to pyrex or aluminum and values of 0.01 correspond to Pt or Ag electrodes.

\subsection{Global model -- partial pressures}
\label{global}

To determine the partial pressures of the background thermal neutral species we applied a global (volume averaged) model of the oxygen 
discharge.  The global model used is discussed in detail by \citet{thorsteinsson10:055008} but 32 additional reactions
have been added to improve the treatment of O$_2$(b$^1\Sigma_{\rm g}^+$), O$_3$, and O$_2^-$ \citep{jonsson18bt}, and in order to make the oxygen reaction set as
detailed as the one discussed by \citet{toneli15:325202}. We explored the partial pressures at 10, 25 and 50 mTorr and the  total absorbed power was  found to be 1.8 W after iteration between the {\tt oopd1} simulations and the 
global model. We assume a cylindrical discharge of diameter 10.25 cm and length 4.5 cm.  We vary the surface quenching coefficient for the singlet metastable molecule O$_2$(a$^1 \Delta_{\rm g})$ in the range $\gamma_{\rm wqa} = 0.00001 - 0.1$ while 
the surface quenching coefficient for the singlet metastable molecule O$_2$(b$^1 \Sigma_{\rm g})$ is kept constant $\gamma_{\rm wqb} = 0.1$.
The fractional densities  found by the global model calculations involving the neutrals O$_2(\mathrm{X}^3 \Sigma_{\rm g}^-)$, O($^3$P), O$_2$(a$^1 \Delta_{\rm g})$ and O$_2$(b$^1 \Sigma_{\rm g})$ are listed in Table \ref{partialpressure}.  These values are used as input for the simulation in the particle-in-cell Monte Carlo collision (PIC/MCC) code {\tt oopd1} as the partial pressure of the neutral background gas.
 Note that not all the neutrals considered in the global model calculations are shown in the table.
We see that the partial pressure of the singlet metastable molecule  O$_2$(a$^1 \Delta_{\rm g})$ increases with decreasing surface 
quenching coefficient and takes its highest value at 25 mTorr for the lowest surface quenching coefficient.  The partial
pressure of  O$_2$(b$^1 \Sigma_{\rm g})$ is always much smaller, maybe due to too large surface quenching coefficient.  
\begin{table}[!t]
\renewcommand\arraystretch{1.5}
		\caption{\label{partialpressure} The partial pressures of the thermal neutrals at 10, 25 and 50 mTorr for different wall quenching coefficients
of the singlet metastable molecule O$_2$(a$^1 \Delta_{\rm g})$ calculated by a global (volume averaged) model. }
		\begin{tabular}{ l  l  l  l  l }
		\hline \hline
    $\gamma_{\rm wqa}$           &	O$_2$$(\mathrm{X}^3 \Sigma_{\rm g}^-)$	&	 O$_2$(a$^1 \Delta_{\rm g})$ & O$_2$(b$^1 \Sigma_{\rm g})$ & O$(^3$P) \\ \hline
10 mTorr                 &                              &			&                &  \\
 $10^{-1}$		&  $0.9926$	& $0.0022$	& $0.0015$ & $0.0015$ \\
			$10^{-2}$		&  $0.9825$	& $0.0124$	& $0.0016$ & $0.0009$ \\
			$10^{-3}$		&  $0.9717$	& $0.0240$	& $0.0018$ & $0.0007$ \\
			$10^{-4}$		&  $0.9684$	& $0.0265$	& $0.0018$ & $0.0015$ \\
			$10^{-5}$		& $0.9681$	& $0.0268$	& $0.0018$	& $0.0015$ \\  \\
25 mTorr   	       &                              &			&                &  \\		
			$10^{-1}$		&  $0.9918$	& $0.0018$	& $0.0012$ & $0.0006$ \\
			$10^{-2}$		&  $0.9847$	& $0.0105$	& $0.0015$ & $0.0007$ \\
			$10^{-3}$		&  $0.9667$	& $0.0290$	& $0.0019$ & $0.0007$ \\
			$10^{-4}$		&  $0.9607$	& $0.0350$	& $0.0019$ & $0.0007$ \\
			$10^{-5}$		& $0.9600$	& $0.0357$	& $0.0020$	& $0.0007$ \\  \\
50 mTorr   	       &                              &			&                &  \\	
	$10^{-1}$		&  $0.9895$	& $0.0013$	& $0.001$ & $0.0003$ \\
			$10^{-2}$		&  $0.9883$	& $0.0054$	& $0.0015$ & $0.0004$ \\
			$10^{-3}$		&  $0.9791$	& $0.0161$	& $0.0020$ & $0.0004$ \\
			$10^{-4}$		&  $0.9739$	& $0.0215$	& $0.0022$ & $0.0004$ \\
			$10^{-5}$		& $0.9732$	& $0.0223$	& $0.0022$	& $0.0004$ \\
\hline \hline
		\end{tabular}
\end{table}

\section{Results and Discussion}
\label{results}

\begin{figure*}
\begin{center}
\includegraphics[scale=0.9]{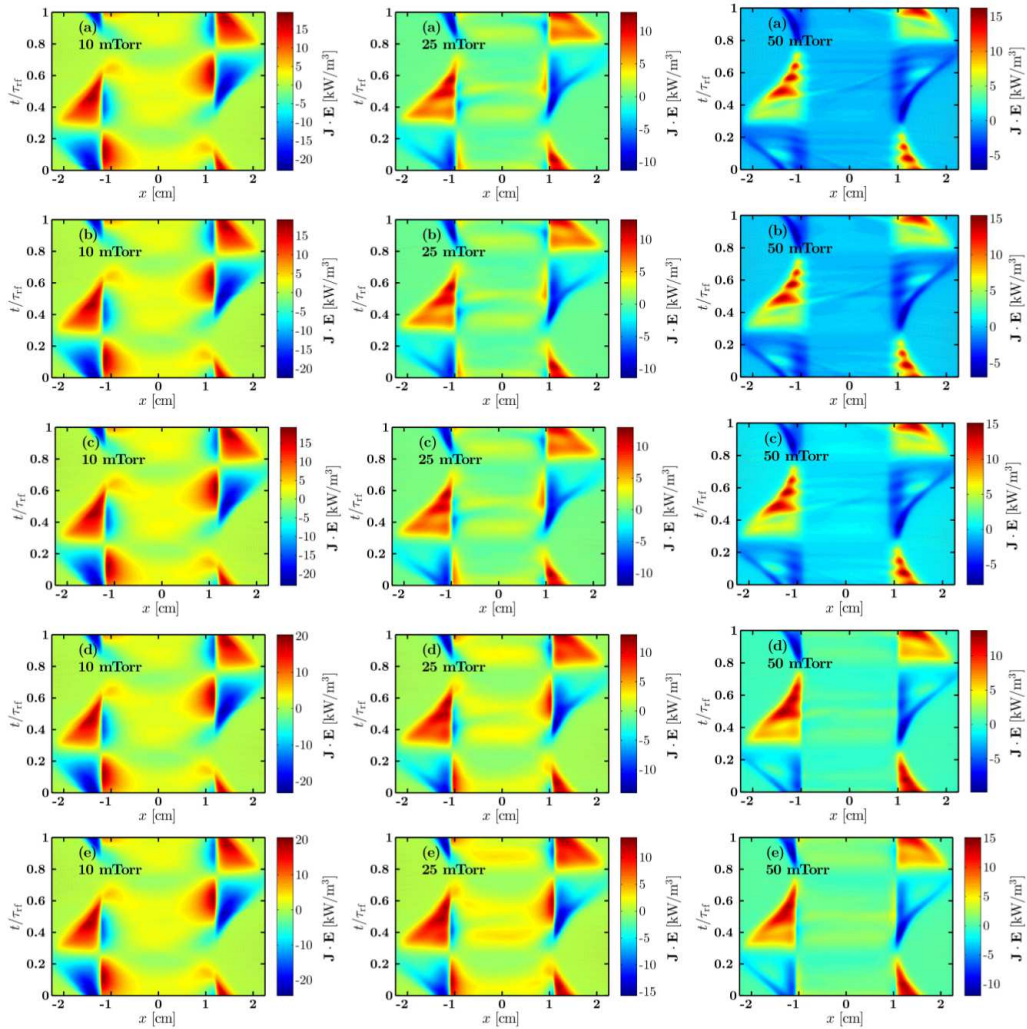}
\end{center}
\caption{\label{spatiotemp10} The spatio-temporal behaviour of the electron power absorption at 10 mTorr (left column), 25 mTorr (center column)  and 50 mTorr (right column) for surface quenching coefficient for the singlet metastable molecule O$_2$(a$^1 \Delta_{\rm g})$ as (a) $\gamma_{\rm wqa}=0.00001$, (b) $\gamma_{\rm wqa}=0.0001$, (c) $\gamma_{\rm wqa}=0.001$, (d) $\gamma_{\rm wqa}=0.01$ and (e) $\gamma_{\rm wqa}=0.1$ for a parallel plate capacitively coupled oxygen discharge with electrode separation of 4.5 cm driven by a 222  V voltage source at driving frequency of 13.56 MHz.}
\end{figure*}

\begin{figure}
\begin{center}
 \includegraphics[scale=0.4]{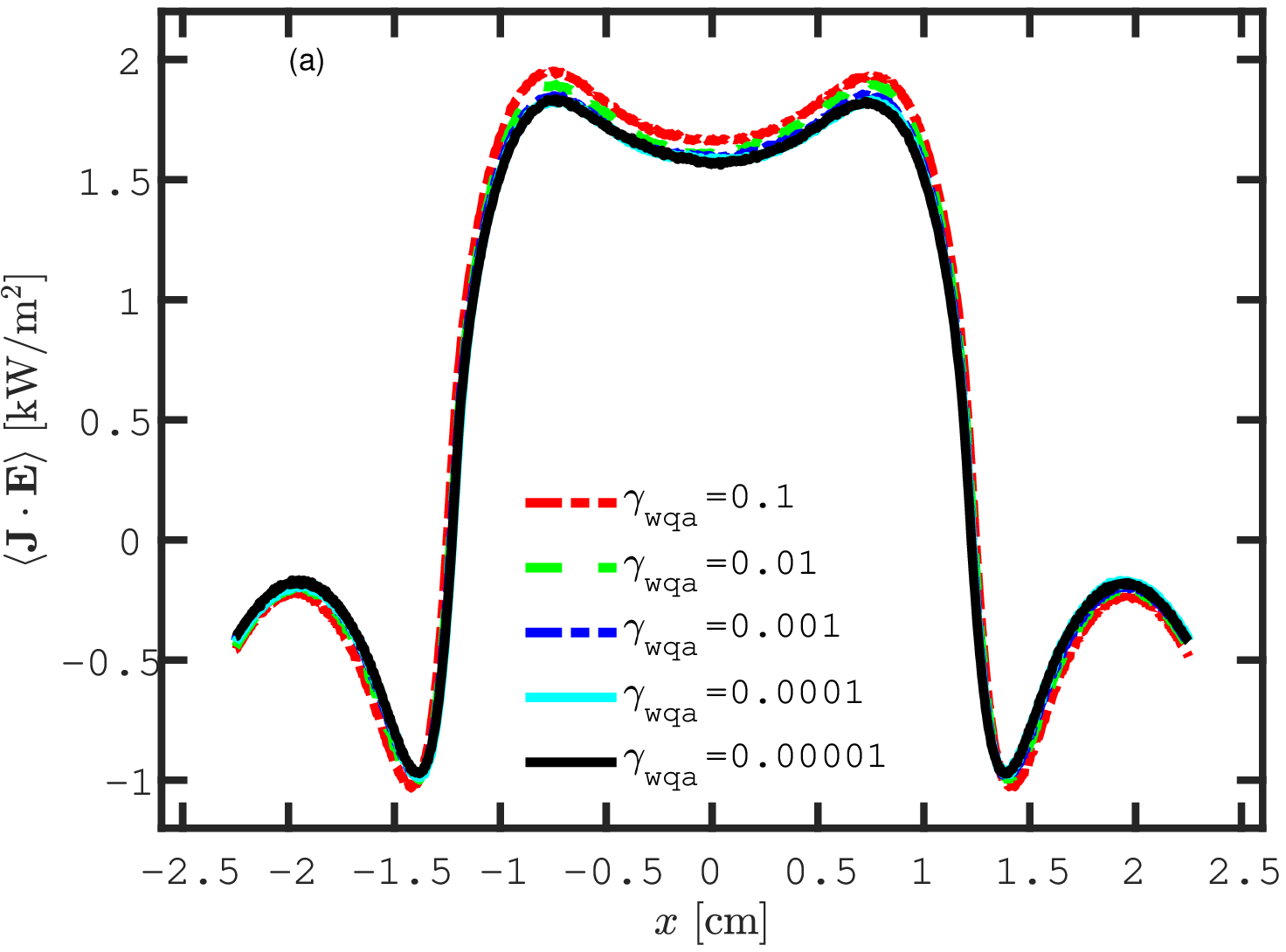}
 
 \includegraphics[scale=0.4]{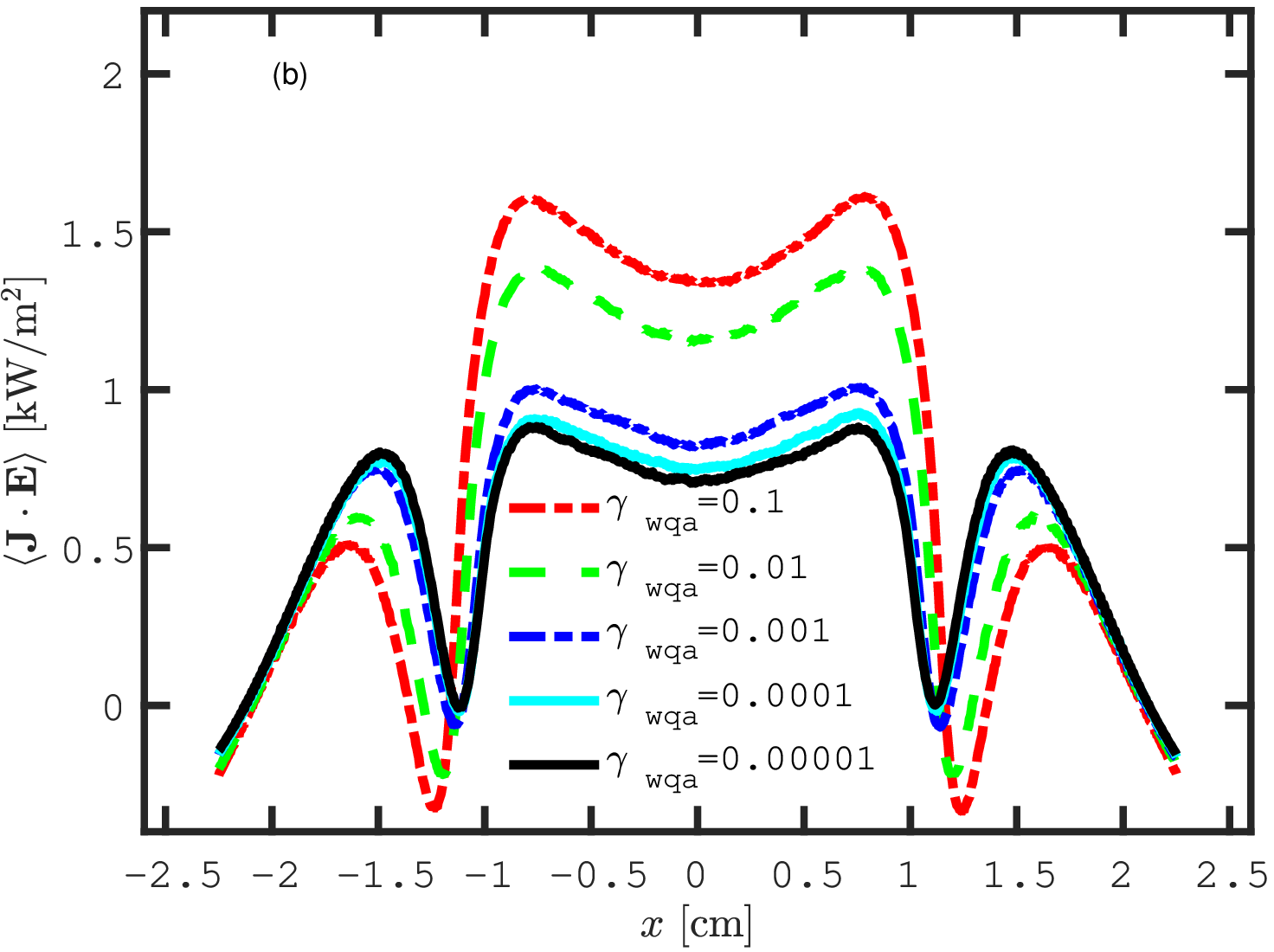}

 \includegraphics[scale=0.4]{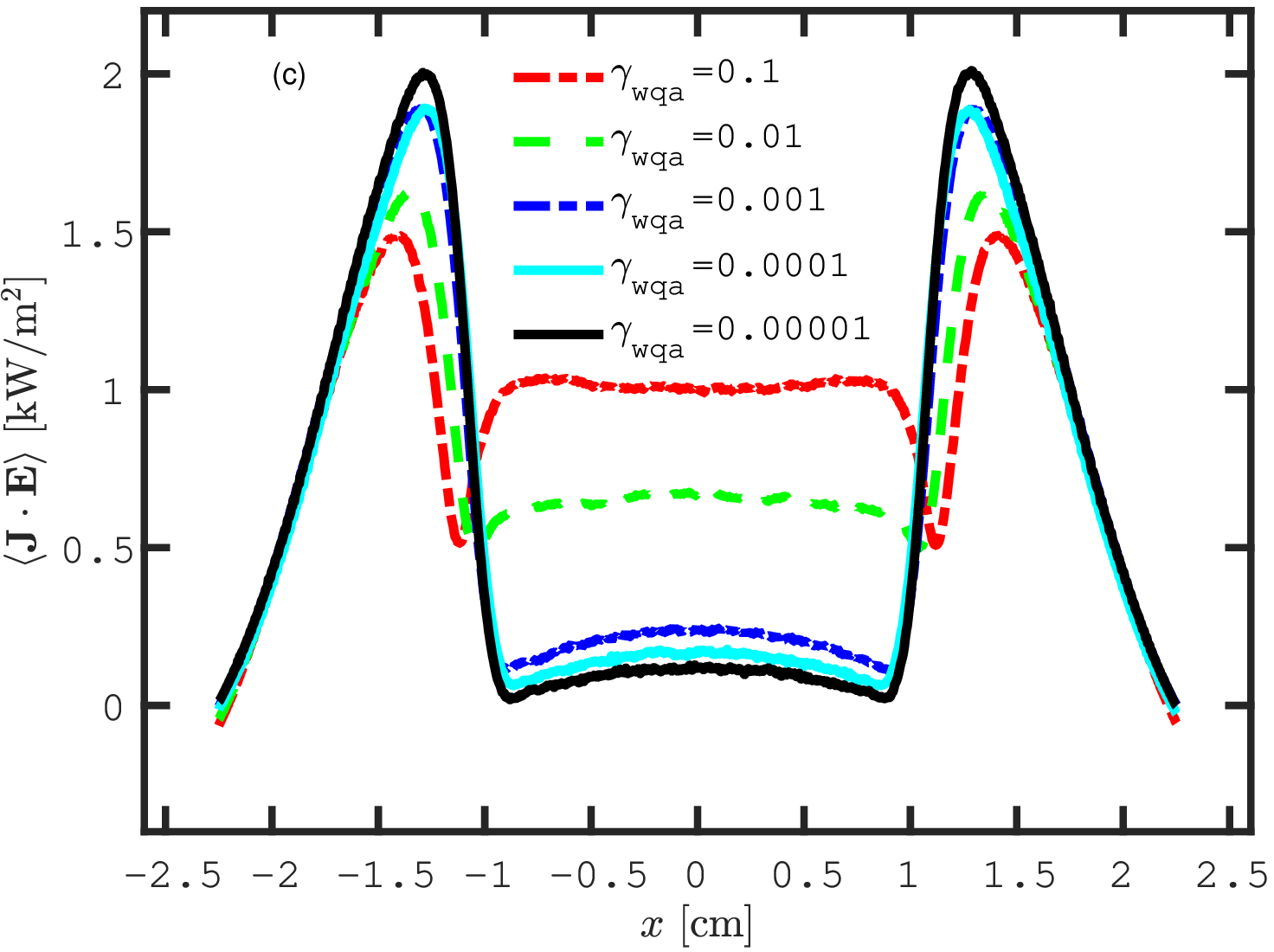}
\end{center}
    \caption{\label{timeavjdote} The time averaged electron heating profile for a parallel plate capacitively coupled oxygen discharge at (a) 10 mTorr, (b) 25 mTorr, and (c) 50  mTorr, with a gap separation of 4.5 cm driven by a 222 V voltage source at driving frequency of 13.56 MHz.}
\end{figure}

Figures \ref{spatiotemp10}
 show the spatio temporal behaviour of the electron power absorption as the surface quenching coefficient for the singlet metastable molecule O$_2$(a$^1 \Delta_{\rm g})$ is varied in the range 0.00001 -- 0.1, for  pressures of 10  mTorr (left column), 25  mTorr (center column) and 50 mTorr (right column).
 The electron power absorption is calculated as
${\bf J}_{\rm e} \cdot {\bf E}$, where ${\bf J}_{\rm e}$ and ${\bf E}$ are the spatially and temporally varying electron current
density and electric field, respectively. 
For each of the figures the abscissa covers the whole inter-electrode gap, from the powered electrode on the left hand side to the grounded electrode on the right hand side. Similarly the ordinate covers the full rf cycle. As displayed in Figure   \ref{spatiotemp10} left column, for low pressure (10 mTorr), the change in the quenching coefficient $\gamma_{\rm wqa}$ does not alter the heating mechanism, which is a combination of a drift ambipolar (DA) heating in the bulk plasma and stochastic heating due to the sheath oscillation ($\alpha$-mode). As the operating pressure is raised to 25  mTorr (see Figure \ref{spatiotemp10} center column)  varying the quenching coefficients clearly has an influence on the heating mechanism.  For a high quenching coefficient the electron heating is a combination of DA- and  $\alpha$-mode, as seen in Figure   \ref{spatiotemp10} center column  (e), similar to what is seen at 10 mTorr independent of the quenching coefficient  (Figure  \ref{spatiotemp10} left column (a) -- (e)).  As the    quenching  coefficient is lowered the bulk heating decreases and stochastic heating in the sheath region becomes more prominent, 
 as seen in Figures \ref{spatiotemp10} center column (d) -- (a). At 50 mTorr   we still see some  bulk heating for the highest quenching coefficients
(Figures \ref{spatiotemp10} right column  (e) -- (d)) but for low  quenching coefficients there is no electron heating observed in the plasma bulk   (Figures \ref{spatiotemp10} right column (c) -- (a)) and pure $\alpha-$mode is observed.  
Figure \ref{timeavjdote} shows the time averaged electron heating profile between the electrodes $\langle {\bf J}_{\rm e} \cdot {\bf E} \rangle$.   We see in Figure \ref{timeavjdote} (a) that at 10 mTorr almost all the electron heating occurs in the plasma bulk (the electronegative core) and the electron heating profile is almost independent of the surface quenching coefficient for the singlet metastable molecule O$_2$(a$^1 \Delta_{\rm g})$. In the sheath regions the time averaged $\langle {\bf J}_{\rm e} \cdot {\bf E} \rangle$ value indicates electron cooling.  This can occur in the sheath regions as the DA-heating  in the bulk maintains the discharge.   At 25 mTorr for high 
 surface quenching coefficient the electron heating in the plasma bulk region dominates as seen in Figure \ref{timeavjdote} (b).  As the  surface quenching coefficient  for the singlet metastable molecule O$_2$(a$^1 \Delta_{\rm g})$ decreases the electron heating in the bulk region decreases and the heating in the sheath regions increases. For surface quenching coefficient in the range 0.001 -- 0.00001 the time averaged heating profile remains roughly the same, electron heating occurs both in the bulk and in the sheath regions, and a combination of DA- and $\alpha-$mode is observed.   When operating at 50 mTorr electron heating in the sheath region dominates as seen in Figure \ref{timeavjdote} (c).  Only for the highest  surface quenching coefficients 0.1 and 0.01 there is some electron heating observed in the bulk region.  For surface quenching coefficients $< 0.001$ there is almost no electron heating in the bulk region at 50 mTorr.  Furthermore, at the higher pressures 25  mTorr and 50  mTorr, the more the quenching coefficient is raised, the more the power absorbtion in the bulk is increased. 
Also, when the power absorbtion in the bulk increases, the power absorbtion in the sheath regions decreases.  
 High frequency oscillations in the electron power absorption
density adjacent to the expanding sheath edge are seen at 25 and 50 mTorr and
 become more clear as the pressure is increased and the 
surface quenching coefficient is decreased.  These oscillations
are a beam-plasma  instability at the electron plasma frequency,  due to
an electron-electron two-stream  instability between the bulk electrons
and electrons accelerated by the moving sheath \citep{oconnell07:034505,wilczek16:063514}. 

\begin{figure}
\begin{center} 
\includegraphics[scale=0.5]{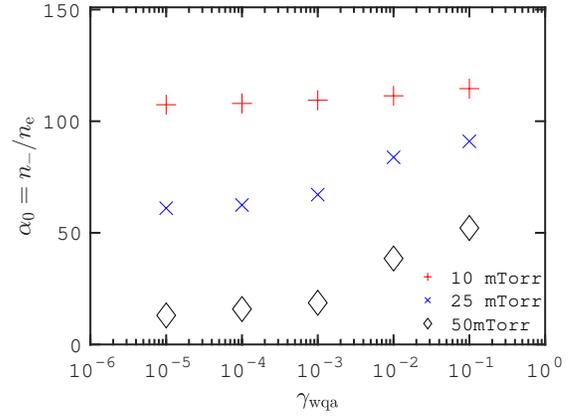}
\end{center} 
\caption{\label{alphavsgamma} The electronegativity in the discharge center ($\alpha_{\rm 0}$) as a function of the quenching coefficient of  the singlet metastable molecule O$_2$(a$^1 \Delta_{\rm g})$   for a parallel plate capacitively coupled oxygen discharge with a gap separation of 4.5 cm driven by a 222 V voltage source at driving frequency of 13.56 MHz.}  
\end{figure}

In order to explore further the observed transition we plot the time averaged center electronegativity 
 as a function of the surface  quenching coefficient for the singlet metastable molecule O$_2$(a$^1 \Delta_{\rm g})$  in Figure \ref{alphavsgamma}. 
   At 10 mTorr the discharge is the most electronegative and the least electronegative at 50 mTorr. 
At 10  mTorr the electronegativity does not  vary much when the surface quenching coefficient is varied. 
The electronegativity is high, in the range 107 -- 114, as the surface quenching coefficient is decreased from 0.1 to
0.00001. 
At 25 and 50 mTorr an increase in the electronegativity is observed with increasing surface quenching coefficient. 
At 25 mTorr the electronegativity increases from 61 to 90, 
and at 50 mTorr the electronegativity increases from 13 to 52 as the  surface quenching coefficient is increased.
 This is related to the fact that, when the surface quenching coefficient is increased, the number of singlet delta  metastable molecules O$_2$(a$^1 \Delta_{\rm g})$ decreases, the negative ion density increases, and the electronegativity increases.  This also means,
as has been pointed out by others  \citep{greb15:044003,derzsi16:015004,derzsi17:034002}, that the surface quenching coefficient
dictates the electronegativity of the oxygen discharge. We note that the electronegativity in this current study is somewhat larger than we have reported in our earlier works \citep{hannesdottir16:055002,gudmundsson18:025009}, particularly at the higher pressures.  This is due to the fact that the density of the singlet metastable molecule O$_2$(a$^1 \Delta_{\rm g})$ is somewhat lower in this current study and the density of the  the singlet metastable molecule   O$_2$(b$^1 \Sigma_{\rm g})$ is significantly lower than assumed in the earlier studies.  These are the results of the improved global model calculations as discussed in section \ref{global}. 
A few measurements of the electronegativity in capacitively 
coupled oxygen discharges   have been reported.  
  \citet{berezhnoj00:800} report a value of around 10 
in a symmetric  capacitively  coupled oxygen discharge with stainless steel electrodes
 operated at 45 mTorr with  electrode spacing of 6 cm.  At 50 mTorr we find  electronegativity of 13
for $\gamma_{\rm wqa} = 10^{-5}$ and 18.7 for $\gamma_{\rm wqa} = 10^{-3}$. Similarly,
 \citet{katsch00:323} estimated the electronegativity in the discharge center of a capacitively 
coupled oxygen discharge, with aluminum electrodes with 2.54 cm separation, to be roughly 2 at 103 mTorr 
and 150 V and to fall below unity at 280 V. We would expect the electronegativity to decrease for 
aluminum electrodes and higher pressure.
For higher pressures \citet{kullig10:065011} reported
electronegativity of 5 -- 6 at 225 mTorr in an asymmetric capacitively coupled discharge with stainless steel electrodes and 
\citet{kaga01:330} measured center electronegativity of 7.4 -- 11.6  at 100 mTorr and 10.8 -- 18.6 at 500 mTorr with aluminum electrodes with spacing of  6 cm.

\begin{figure}
\begin{center} 
    \includegraphics[scale=0.4]{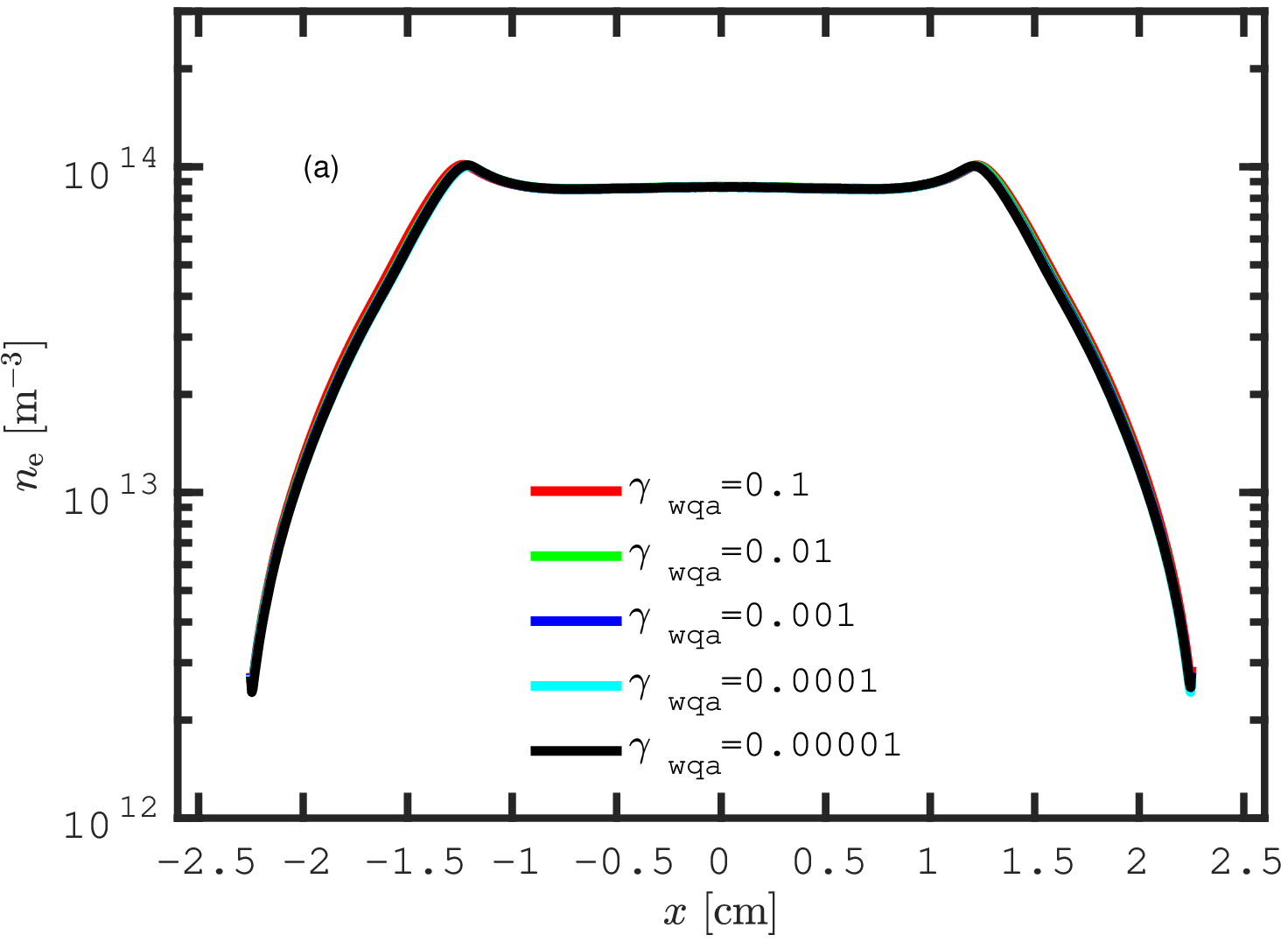}

    \includegraphics[scale=0.4]{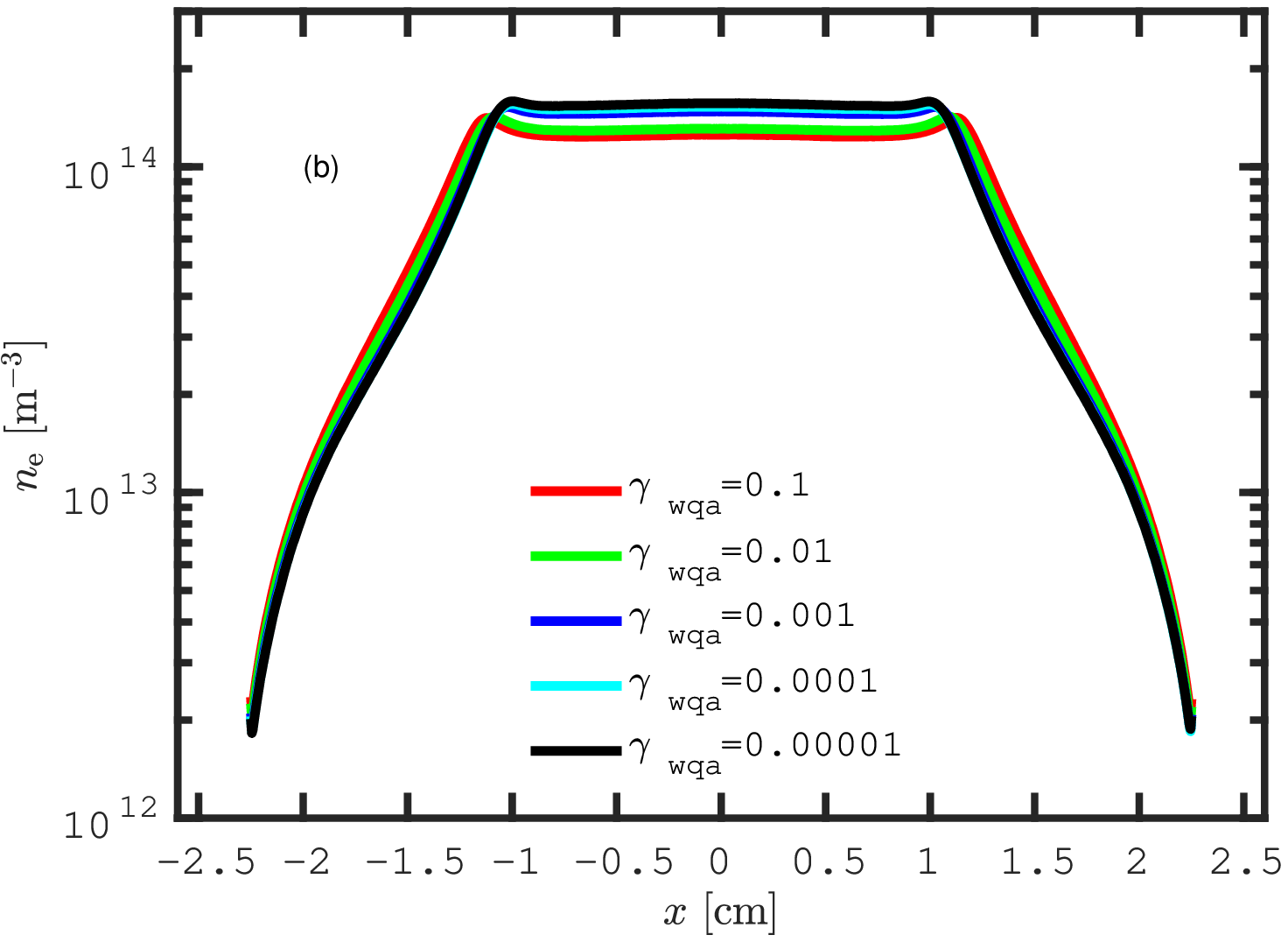}

    \includegraphics[scale=0.4]{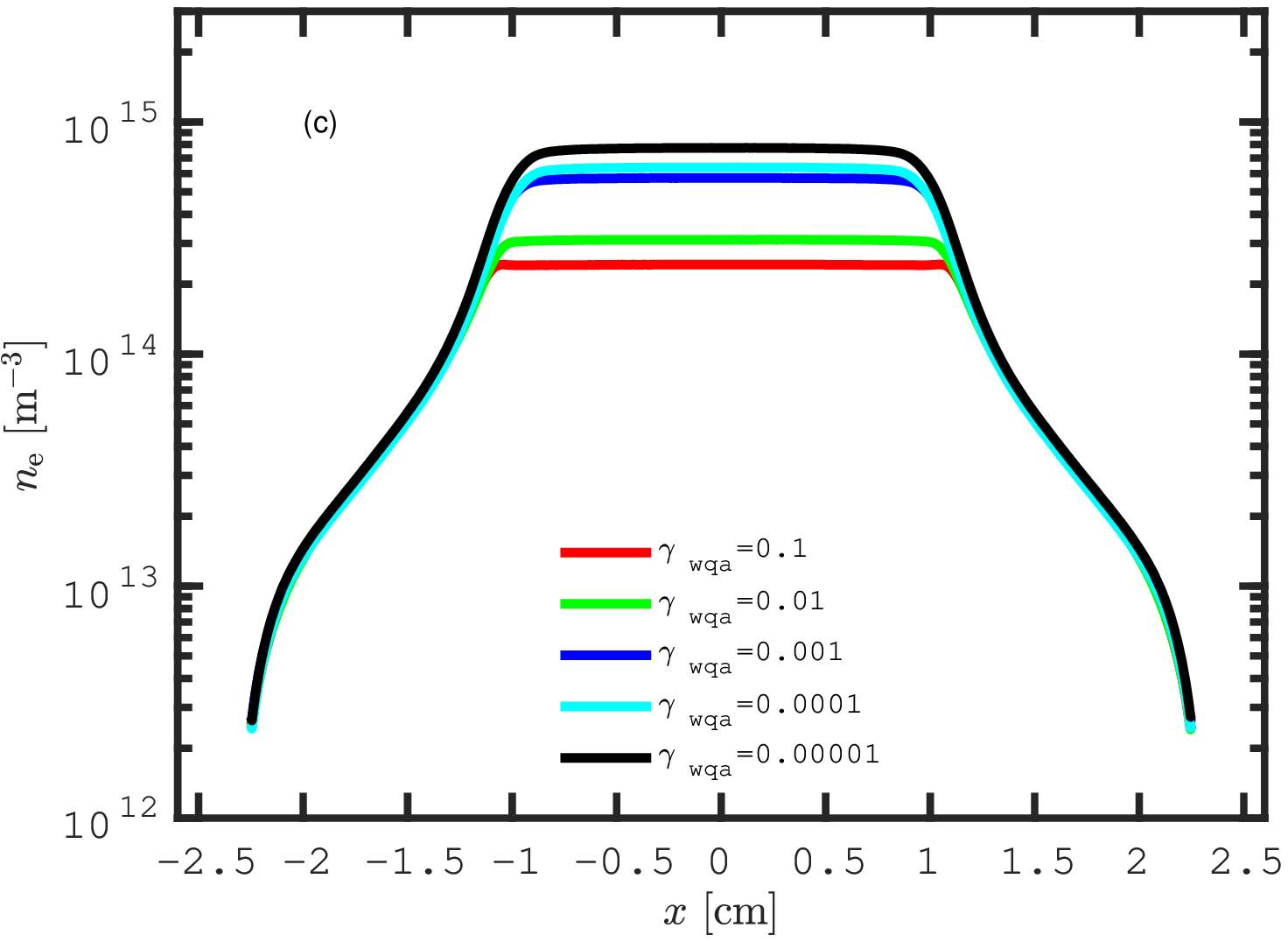}
\end{center} 
  \caption{\label{eldens} The electron density  profile for a parallel plate capacitively coupled oxygen discharge at (a) 10  mTorr, (b) 25 mTorr, and (c) 50 mTorr,  with a gap separation of 4.5  cm driven by a 222  V voltage source at driving frequency of 13.56  MHz.}
  \end{figure}

  The time averaged electron density profile is shown in Figure \ref{eldens}.   The center electron density at 10 mTorr is $8.6 \times 10^{13}$ m$^{-3}$ and is roughly independent of the surface quenching coefficient as seen in Figure \ref{eldens} (a).  The electron density profile is flat within the plasma bulk with a peak close to the sheath edge.  At 25 mTorr the center electron density is  $1.25 \times 10^{14}$ m$^{-3}$ for $\gamma_{\rm wqa} = 0.1$ and increases slightly with decreasing quenching coefficient or to  $1.57 \times 10^{14}$ m$^{-3}$ for $\gamma_{\rm wqa} = 0.00001$.   The electron density profile is flat within the plasma bulk and an increase in the density is observed close to the sheath edge. 
 At 50 mTorr the electron density is  $2.4 \times 10^{14}$ m$^{-3}$ for $\gamma_{\rm wqa} = 0.1$ and increases with decreasing quenching coefficient to  $7.7 \times 10^{14}$ m$^{-3}$ for $\gamma_{\rm wqa} = 0.00001$.  The electron density profile is flat within the plasma bulk and here no increase (or a peak) is observed at the sheath edge. 
\citet{kechkar17:065009} explored experimentally slightly asymmetric capacitively couple oxygen discarge operated at 13.56 MHz,  and  report electron density in the range 10$^{15}$ -- $7 \times  10^{16}$ m$^{-3}$ for a discharge operated at 100 mTorr while  the power is varied in the range 30 -- 600 W.
When operating  low power of 30 W, in what they refer to as  the $\alpha-$mode, the electron density is in the range $6  \times  10^{14}$ -- $1.6  \times  10^{15}$ m$^{-3}$, increasing with increasing pressure from 10 -- 50 mTorr. These values measured at low 
power of 30 W and in the pressure range 10 -- 50 mTorr
are very similar to the simulated electron density  values reported here. Also  \citet{katsch00:323} report eletron density of    $3  \times  10^{14}$   m$^{-3}$ at 150 V and $1.4  \times  10^{15}$ at 280 V when operating oxygen discharge at 100 mTorr and   \citet{berezhnoj00:800} report electron density of    $1.1  \times  10^{14}$   m$^{-3}$
in a symmetric  capacitively  coupled oxygen  discharge  with stainless steel electrodes operated at 45 mTorr with  electrode spacing of 6 cm,  
and current density of $J = 0.31$ mA/cm$^2$. 

\begin{figure}
\begin{center} 
   \includegraphics[scale=0.45]{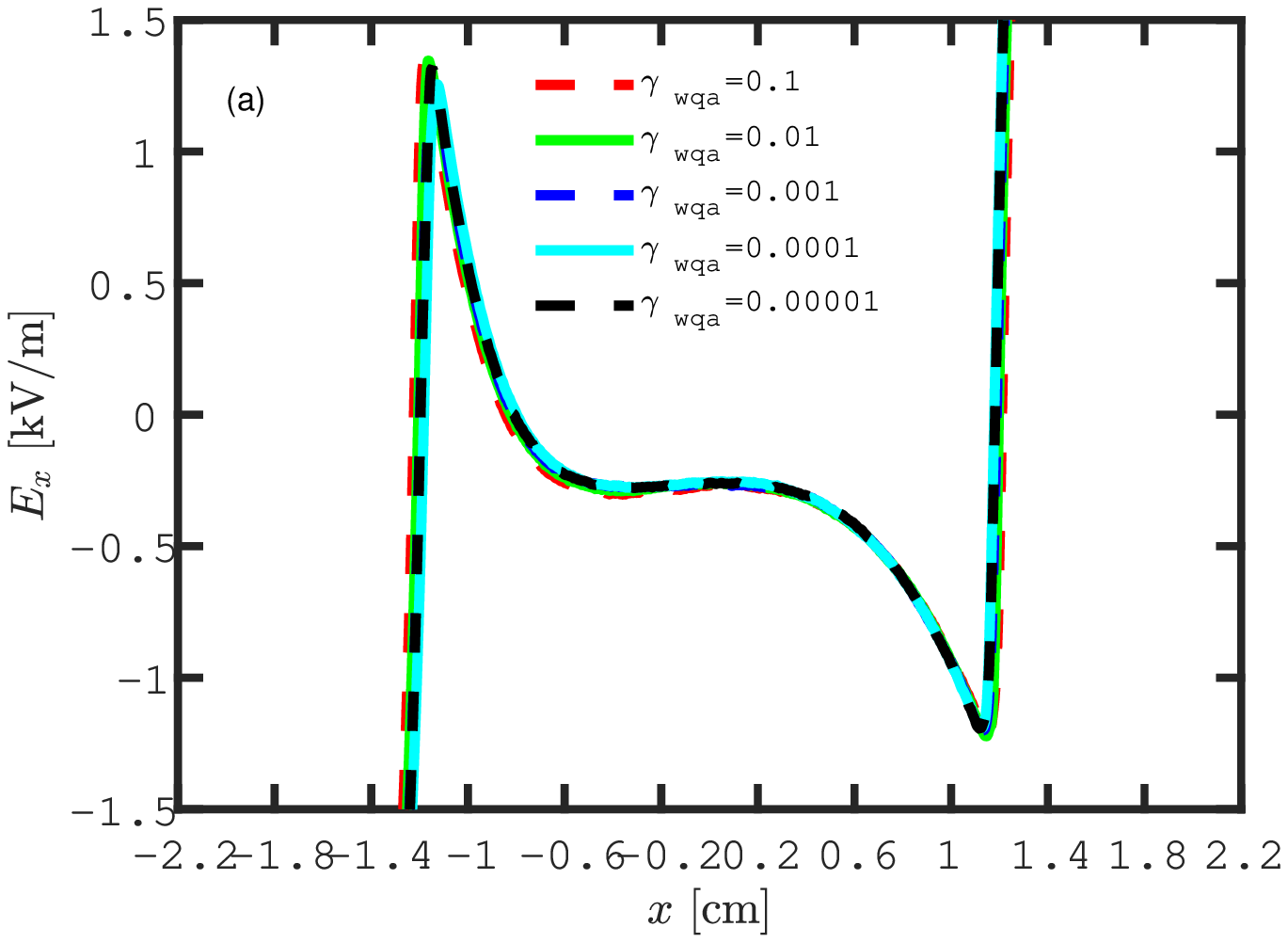}

   \includegraphics[scale=0.45]{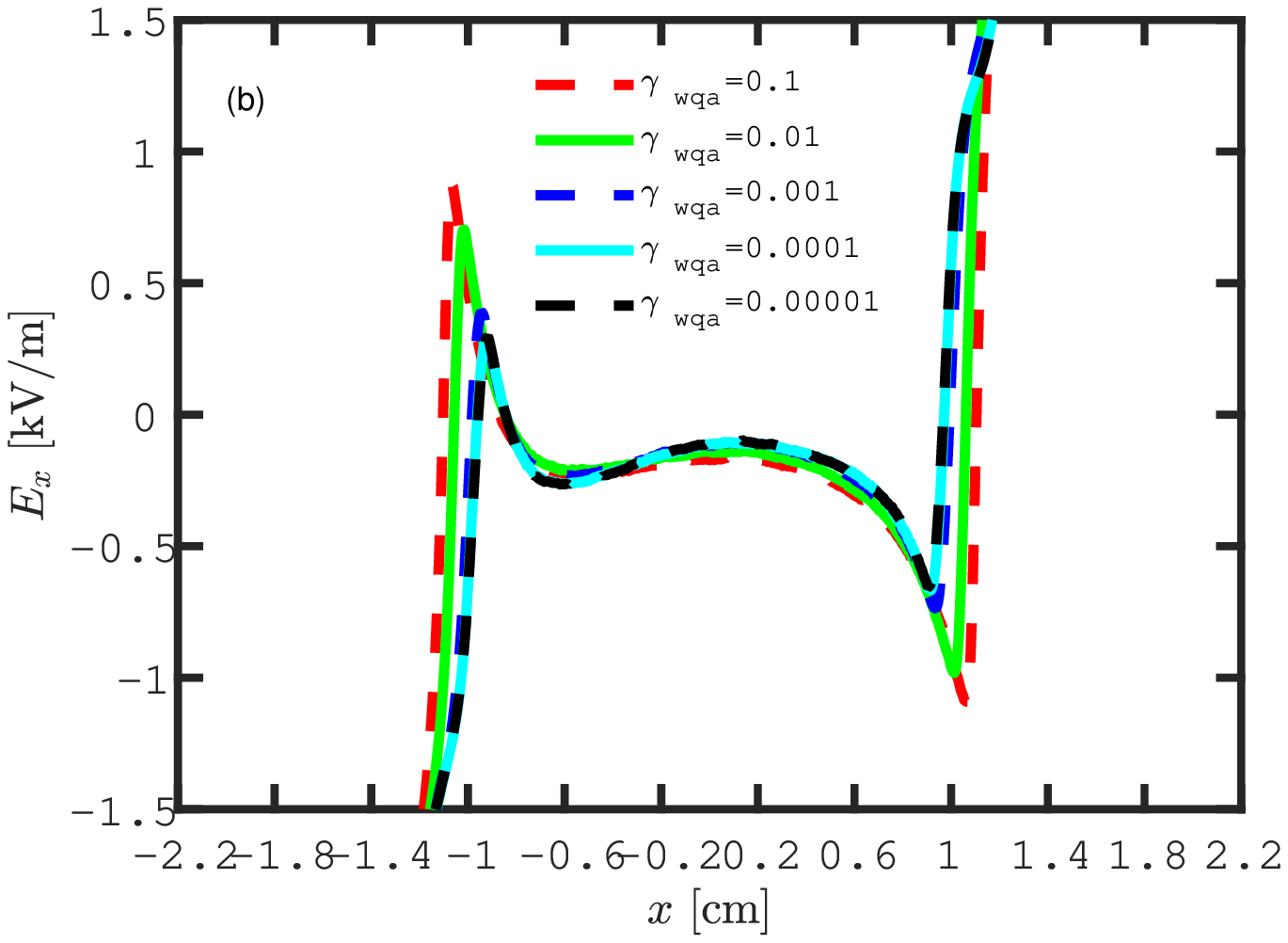}

   \includegraphics[scale=0.45]{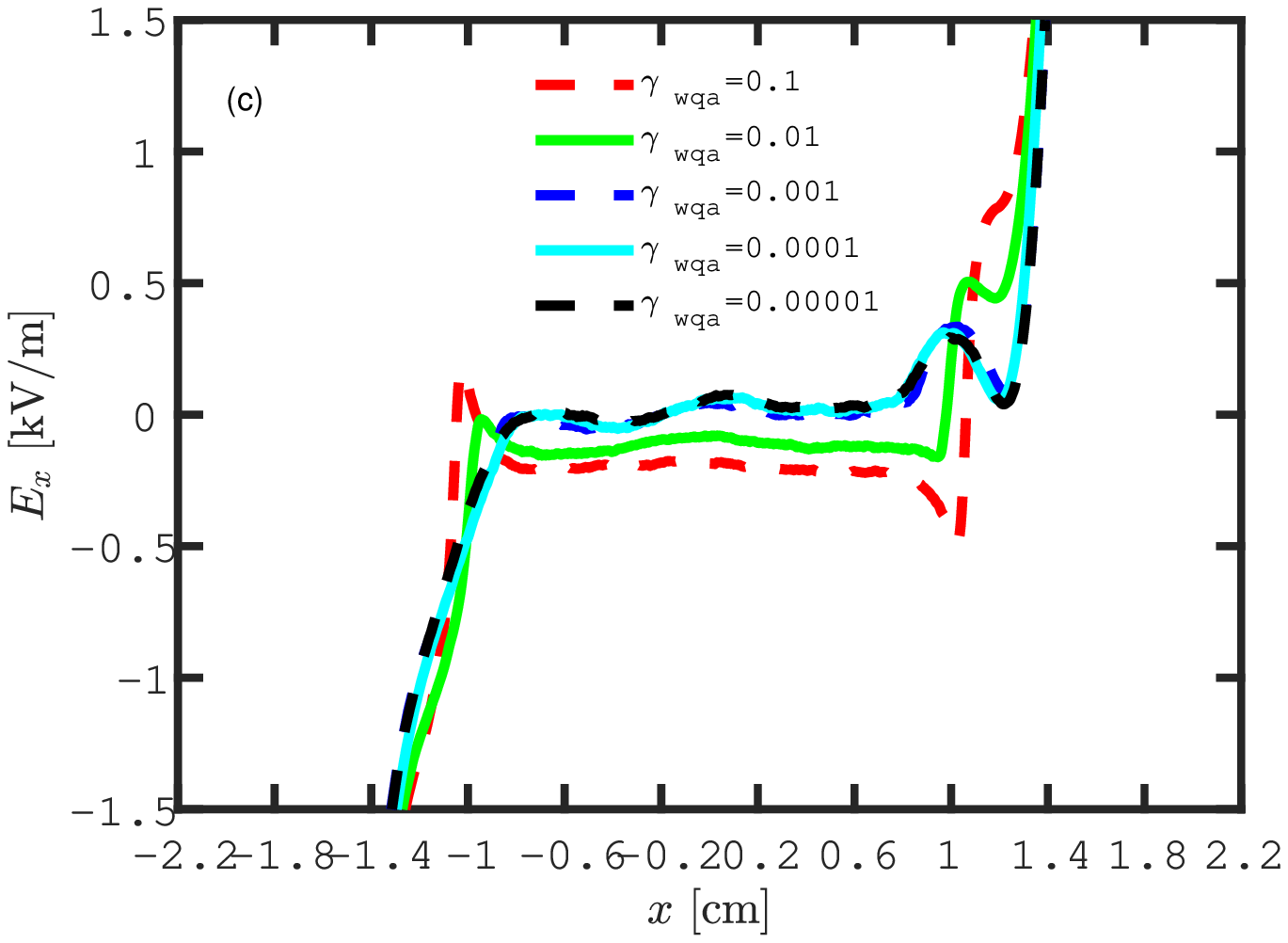}
\end{center} 
    \caption{\label{efieldx} The electric field in the bulk region at $t/\tau_{rf} = 0.5$  for a parallel plate capacitively coupled oxygen discharge at (a) 10  mTorr, (b) 25 mTorr, and (c) 50 mTorr with a gap separation of 4.5 cm driven by a 222 V voltage source at driving frequency of 13.56 MHz.}
\end{figure}

Figure \ref{efieldx} shows the electric field profile in the plasma bulk for the various  quenching coefficients  at a fixed time slice $t/\tau_{\rm rf} = 0.5$.  Note that these curves are not time averaged.  At 10 mTorr there is a strong electric field gardient in the bulk region and the electric field strength and profile is independent of the surface quenching coefficient as seen in Figure \ref{efieldx} (a).   
  The electric field is almost flat and takes its lowest absolute value in the center of the  electronegative core, while it assumes strong values as the sheath region is approached.  The shape of the electric field profile is similar to the one
predicted by the simple model of \citet{schulze11:275001}. The resulting electric field is a combination of a drift field and an ambipolar field.  The drift electric field is due to low bulk conductivity or low electron density.  We have seen in Figure \ref{eldens} that the electron density is very low indeed. The peak in the electric field at the sheath edge is mainly
caused by a local maximum of the electron density at the sheath edge and the corresponding
high value of $\partial n_{\rm e}/ \partial x$ on the plasma bulk side of this maximum.
At this location  diffusion directs the electrons into the plasma bulk, while positive ions 
flow continuously   toward the electrode. This generates an ambipolar field, that couples electron and positive 
ion motion and accelerates electrons towards the electrode.  
At 25 mTorr and 50 mTorr (Figure \ref{efieldx} (b) and Figure \ref{efieldx}  (c)) important changes are observed.  For both the pressures considered, the higher the  quenching coefficient, the higher is the electric field peak in the sheath region. 
There is clearly a transition from DA-$\alpha$-mode to $\alpha-$mode when increasing the operating pressure from 25 mTorr to 50 mTorr. At 50 mTorr and low quenching coefficient the electric field is flat and no peaks are observed on the plasma bulk side of the sheath edge.
Transitions  between  the DA-mode and the $\alpha$-mode have been demonstrated by both simulations 
and experiments on  CF$_4$ discharges \citep{schulze11:275001,liu15:034006} where by
 increasing the operating pressure at a fixed applied 
voltage,  a  transition  from  the  $\alpha$-mode  to  the  DA-mode  is 
induced.  Note that the CF$_4$ discharge is weakly electronegative
at 75 mTorr while it is strongly electronegative at 600 mTorr \citep{derzsi15:346}.  
Also by increasing  the  voltage  at  a fixed  pressure,  a transition from the DA-mode 
to the $\alpha$-mode is observed  in a CF$_4$ discharge \citep{schulze11:275001}.
Oxygen behaves in the opposite way,  by increasing the pressure at a given voltage 
a transition from the DA-$\alpha$-mode to the $\alpha$-mode is observed in the oxygen discharge \citep{gudmundsson17:193302}. 
This is a similar to the transition  reported by  Derzsi et al.~\cite{derzsi17:034002} 
which observe an  operation mode transition from DA-$\alpha$-mode to $\alpha$-mode in an oxygen discharge 
as harmonics are added to the voltage waveforms for 10 and 15 MHz driving frequency,
which also  coincides with a strong decrease in the electronegativity.

\begin{figure}
\begin{center} 
    \includegraphics[scale=0.55]{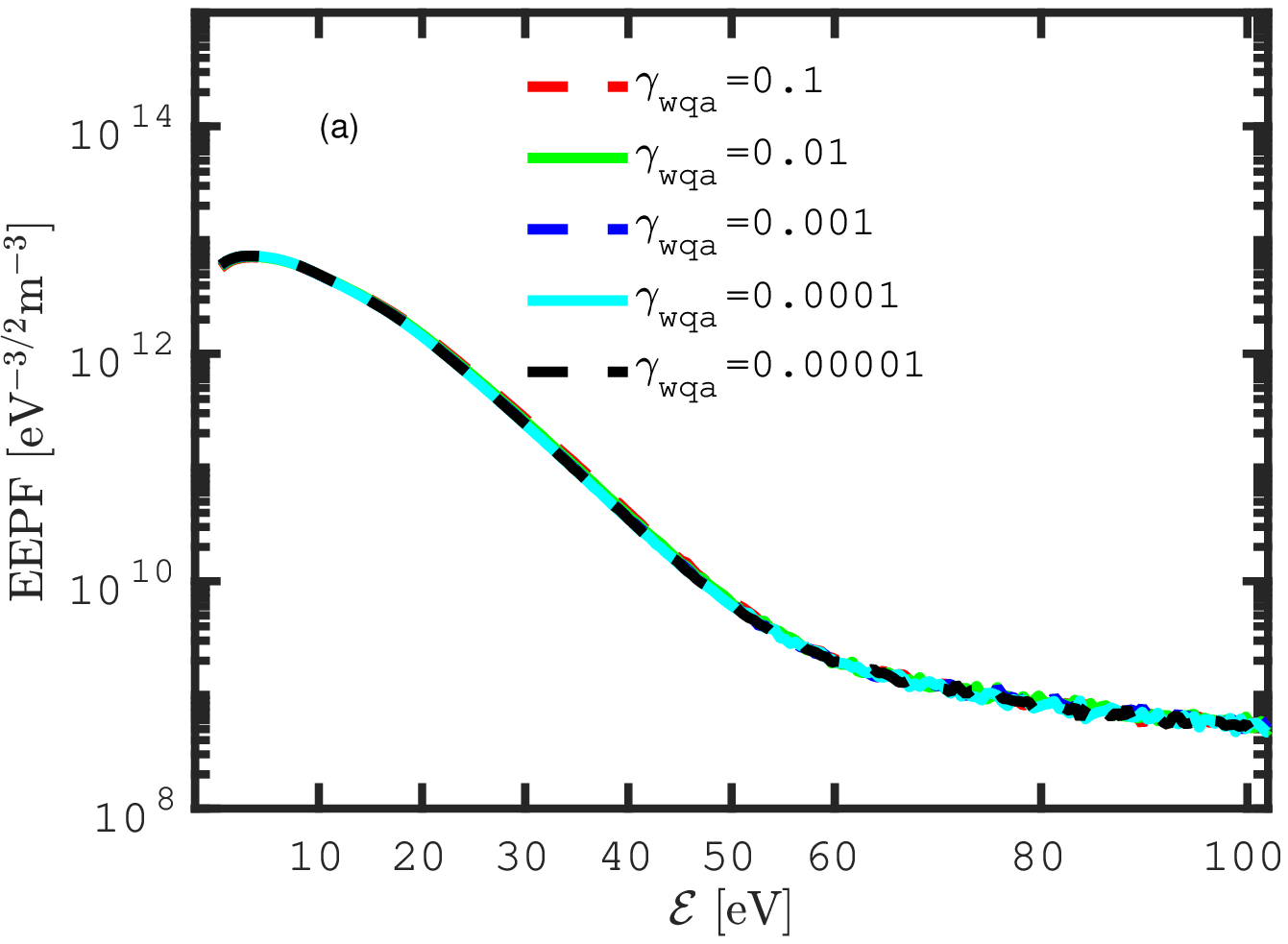}

    \includegraphics[scale=0.55]{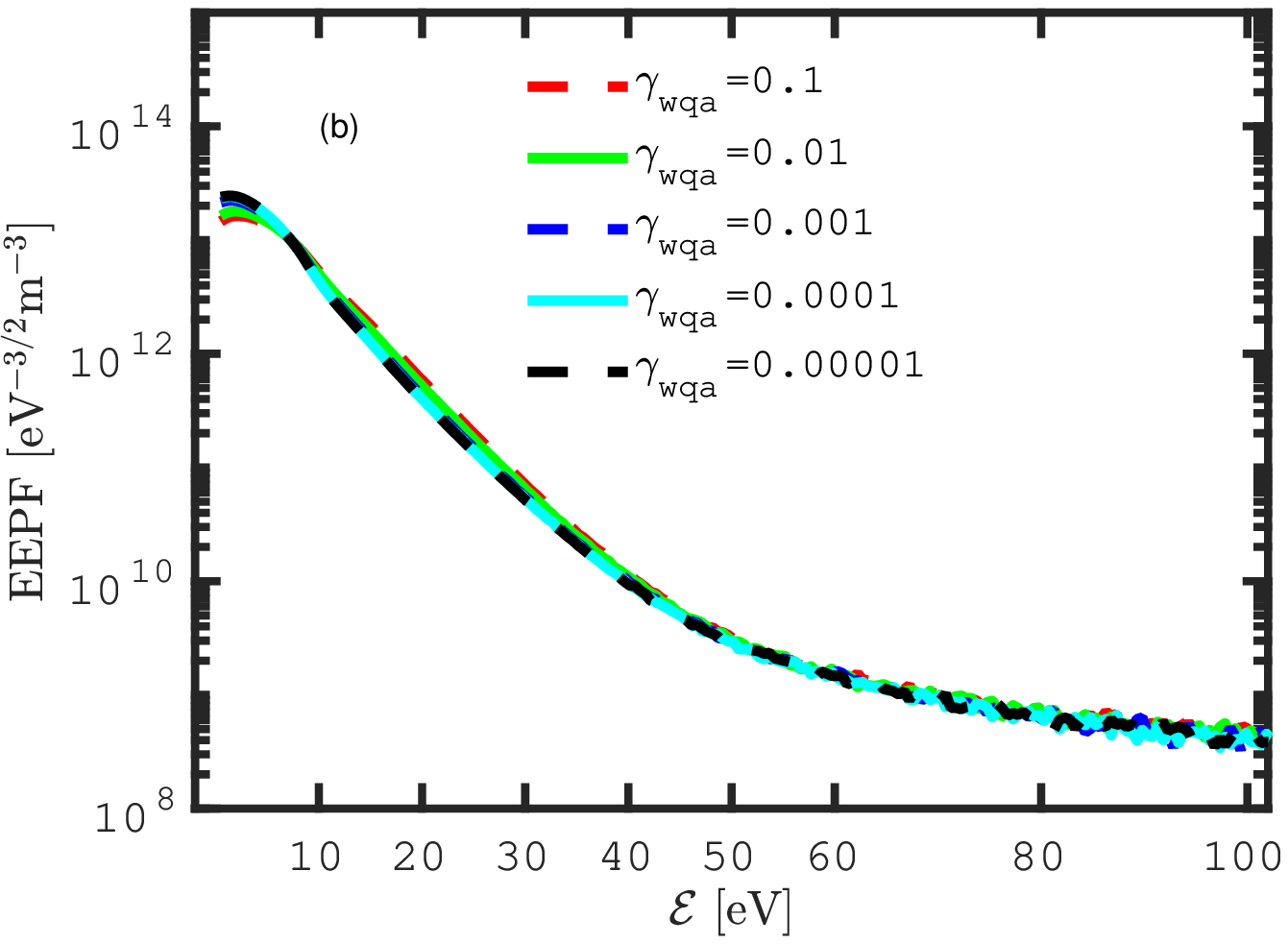}

    \includegraphics[scale=0.55]{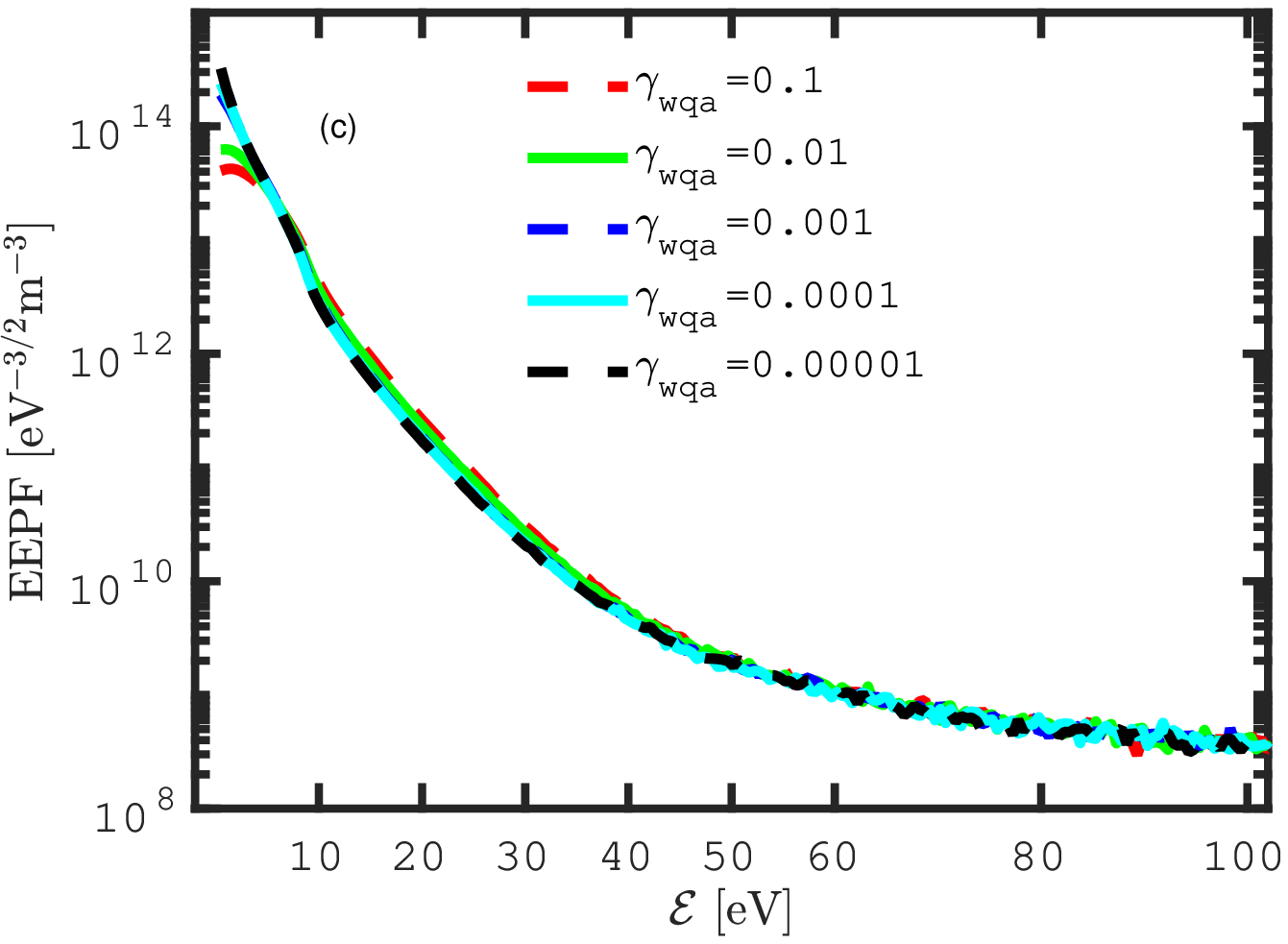}
\end{center} 
\caption{\label{eepf} The electron energy probability function (EEPF) in the discharge center for a parallel plate capacitively coupled oxygen discharge at (a) 10  mTorr, (b) 25 mTorr, and (c) 50 mTorr with a gap separation of 4.5  cm driven by a 222  V voltage source at driving frequency of 13.56  MHz.}
\end{figure}
The evolution of the electron energy probability function (EEPF) in the discharge center with surface quenching coefficient $\gamma_{\rm wqa}$ is shown in
Figure \ref{eepf}. At the lower pressures 10 mTorr and 25  mTorr (Figures  \ref{eepf} (a) and (b)) the EEPF  curves outward for every quenching coefficient value. This is due to the significant contribution of the DA-mode to the bulk electron heating as clearly seen in Figures \ref{timeavjdote} (a) and (b) for 10 and 25 mTorr, respecitvely.  At the higher pressure of 50 mTorr the EEPF  still { curves outward for the highest  quenching coefficients (as there is DA-heating present)  but it transitions  to  curve inward   (bi-Maxwellian) for the lowest quenching coefficients. 
The  bi-Maxwellian shape of the  EEPF in CCPs is commonly associated with predominant sheath heating ($\alpha$-mode). The low energy electron population represents electrons confined in the bulk plasma by an ambipolar potential, which are only weakly heated by the rf field, while the high energy population participates in the sheath heating.
So the population of low energy electrons is high, as the bulk heating mechanism is weak. At 50  mTorr the electon probability function shows the highest value for low surface quenching, i.e.~it has transitioned to become a bi-Maxwellian, that is when  the sheath heating mechanism predominates, as was seen in Figure \ref{spatiotemp10} right column (a) -- (c), where the bulk heating is almost absent.


\section{Conclusion}
\label{conclusion}

The one-dimensional object-oriented particle-in-cell Monte Carlo collision code {\tt oopd1}
was applied  to explore the  evolution of the electron heating mechanism and the EEPF 
in  a capacitively coupled oxygen discharge while the wall quenching probability of 
the single metastable molecule O$_2$(a$^1\Delta_{\rm g}$) is varied.
We find that at low pressure (10 mTorr) the surface quenching coefficient has no influence on the 
electron heating mechanism and electron heating is dominated by drift-ambipolar heating in the plasma bulk
and electron cooling is observed in the sheath region.  At 25 mTorr the electron heating
exhibits a  combination of DA-mode and $\alpha-$mode.  For the highest quenching coefficient the DA-mode dominates, 
but the role of the DA-mode decreases with decreasing quenching coefficient.  At the highest pressure explored,
 50 mTorr,  electron heating  in the sheath region dominates. However, for the highest quenching coefficient there is some contribution from the 
DA-mode in the plasma bulk, but this contribution decreases to almost zero, and thus a pure $\alpha-$mode is observed
 for quenching coefficient of 0.001 or smaller.
We have demonstrated that the surface quenching coefficient of the singlet metastable molecule  O$_2$(a$^1\Delta_{\rm g}$), and thus the electrode material, 
more or less dictates the electronegativity within the discharge and the electron heating mechanism, except at very low operating pressure ($\sim$ 10 mTorr).  However, the quenching coefficients, even for the most common electrodes, are not very well known.  

\acknowledgements

The authors are thankful to Ragnar D.~B.~J{\'o}nsson for assistance with the global model calculations.
This work was partially supported by the  Icelandic Research Fund Grant
No.~163086, the University of Iceland Research Fund, 
  and the Swedish Government Agency for Innovation Systems (VINNOVA) contract no. 2014-04876. 



\begin{thebibliography}{70}
\expandafter\ifx\csname natexlab\endcsname\relax\def\natexlab#1{#1}\fi
\expandafter\ifx\csname bibnamefont\endcsname\relax
  \def\bibnamefont#1{#1}\fi
\expandafter\ifx\csname bibfnamefont\endcsname\relax
  \def\bibfnamefont#1{#1}\fi
\expandafter\ifx\csname citenamefont\endcsname\relax
  \def\citenamefont#1{#1}\fi
\expandafter\ifx\csname url\endcsname\relax
  \def\url#1{\texttt{#1}}\fi
\expandafter\ifx\csname urlprefix\endcsname\relax\def\urlprefix{URL }\fi
\providecommand{\bibinfo}[2]{#2}
\providecommand{\eprint}[2][]{\url{#2}}

\bibitem[{\citenamefont{Lieberman and Godyak}(1998)}]{lieberman98:955}
\bibinfo{author}{\bibfnamefont{M.~A.}~\bibnamefont{Lieberman}} \bibnamefont{and}
  \bibinfo{author}{\bibfnamefont{V.}~\bibnamefont{Godyak}},
  \bibinfo{journal}{IEEE Transactions on Plasma Science}
  \textbf{\bibinfo{volume}{26}}, \bibinfo{pages}{955} (\bibinfo{year}{1998}).

\bibitem[{\citenamefont{Gozadinos et~al.}(2001)\citenamefont{Gozadinos, Vender,
  Turner, and Lieberman}}]{gozadinos01:117}
\bibinfo{author}{\bibfnamefont{G.}~\bibnamefont{Gozadinos}},
  \bibinfo{author}{\bibfnamefont{D.}~\bibnamefont{Vender}},
  \bibinfo{author}{\bibfnamefont{M.~M.} \bibnamefont{Turner}},
  \bibnamefont{and} \bibinfo{author}{\bibfnamefont{M.~A.}
  \bibnamefont{Lieberman}}, \bibinfo{journal}{Plasma Sources Science and
  Technology} \textbf{\bibinfo{volume}{10}}, \bibinfo{pages}{117}
  (\bibinfo{year}{2001}).

\bibitem[{\citenamefont{Wood}(1991)}]{wood91t}
\bibinfo{author}{\bibfnamefont{B.~P.} \bibnamefont{Wood}}, Ph.D. thesis,
  \bibinfo{school}{University of California at Berkeley}
  (\bibinfo{year}{1991}).

\bibitem[{\citenamefont{Wood et~al.}(1995)\citenamefont{Wood, Lieberman, and
  Lichtenberg}}]{wood95:89}
\bibinfo{author}{\bibfnamefont{B.~P.} \bibnamefont{Wood}},
  \bibinfo{author}{\bibfnamefont{M.~A.} \bibnamefont{Lieberman}},
  \bibnamefont{and} \bibinfo{author}{\bibfnamefont{A.~J.}
  \bibnamefont{Lichtenberg}}, \bibinfo{journal}{IEEE Transactions on Plasma
  Science} \textbf{\bibinfo{volume}{23}}, \bibinfo{pages}{89}
  (\bibinfo{year}{1995}).

\bibitem[{\citenamefont{Liu et~al.}(2011)\citenamefont{Liu, Zhang, Jiang, Hou,
  Jiang, Lu, and Wang}}]{liu11:055002}
\bibinfo{author}{\bibfnamefont{Y.-X.} \bibnamefont{Liu}},
  \bibinfo{author}{\bibfnamefont{Q.-Z.} \bibnamefont{Zhang}},
  \bibinfo{author}{\bibfnamefont{W.}~\bibnamefont{Jiang}},
  \bibinfo{author}{\bibfnamefont{L.-J.} \bibnamefont{Hou}},
  \bibinfo{author}{\bibfnamefont{X.-Z.} \bibnamefont{Jiang}},
  \bibinfo{author}{\bibfnamefont{W.-Q.} \bibnamefont{Lu}}, \bibnamefont{and}
  \bibinfo{author}{\bibfnamefont{Y.-N.} \bibnamefont{Wang}},
  \bibinfo{journal}{Physical Review Letters} \textbf{\bibinfo{volume}{107}},
  \bibinfo{pages}{055002} (\bibinfo{year}{2011}).

\bibitem[{\citenamefont{Liu et~al.}(2012)\citenamefont{Liu, Zhang, Liu, Song,
  Bogaerts, and Wang}}]{liu12:114101}
\bibinfo{author}{\bibfnamefont{Y.-X.} \bibnamefont{Liu}},
  \bibinfo{author}{\bibfnamefont{Q.-Z.} \bibnamefont{Zhang}},
  \bibinfo{author}{\bibfnamefont{J.}~\bibnamefont{Liu}},
  \bibinfo{author}{\bibfnamefont{Y.-H.} \bibnamefont{Song}},
  \bibinfo{author}{\bibfnamefont{A.}~\bibnamefont{Bogaerts}}, \bibnamefont{and}
  \bibinfo{author}{\bibfnamefont{Y.-N.} \bibnamefont{Wang}},
  \bibinfo{journal}{Applied Physics Letters} \textbf{\bibinfo{volume}{101}},
  \bibinfo{eid}{114101} (\bibinfo{year}{2012}).

\bibitem[{\citenamefont{Wilczek et~al.}(2015)\citenamefont{Wilczek,
  Trieschmann, Schulze, Sch{\"u}ngel, Brinkmann, Derzsi, Korolov, Donk{\'o},
  and Mussenbrock}}]{wilczek15:024002}
\bibinfo{author}{\bibfnamefont{S.}~\bibnamefont{Wilczek}},
  \bibinfo{author}{\bibfnamefont{J.}~\bibnamefont{Trieschmann}},
  \bibinfo{author}{\bibfnamefont{J.}~\bibnamefont{Schulze}},
  \bibinfo{author}{\bibfnamefont{E.}~\bibnamefont{Sch{\"u}ngel}},
  \bibinfo{author}{\bibfnamefont{R.~P.} \bibnamefont{Brinkmann}},
  \bibinfo{author}{\bibfnamefont{A.}~\bibnamefont{Derzsi}},
  \bibinfo{author}{\bibfnamefont{I.}~\bibnamefont{Korolov}},
  \bibinfo{author}{\bibfnamefont{Z.}~\bibnamefont{Donk{\'o}}},
  \bibnamefont{and}
  \bibinfo{author}{\bibfnamefont{T.}~\bibnamefont{Mussenbrock}},
  \bibinfo{journal}{Plasma Sources Science and Technology}
  \textbf{\bibinfo{volume}{24}}, \bibinfo{pages}{024002}
  (\bibinfo{year}{2015}).

\bibitem[{\citenamefont{Czarnetzki et~al.}(2006)\citenamefont{Czarnetzki,
  Mussenbrock, and Brinkmann}}]{czarnetzki06:123503}
\bibinfo{author}{\bibfnamefont{U.}~\bibnamefont{Czarnetzki}},
  \bibinfo{author}{\bibfnamefont{T.}~\bibnamefont{Mussenbrock}},
  \bibnamefont{and}
  \bibinfo{author}{\bibfnamefont{R.}~\bibnamefont{Brinkmann}},
  \bibinfo{journal}{Physics of Plasmas} \textbf{\bibinfo{volume}{13}},
  \bibinfo{pages}{123503} (\bibinfo{year}{2006}).

\bibitem[{\citenamefont{Donk\'{o} et~al.}(2009)\citenamefont{Donk\'{o},
  Schulze, Czarnetzki, and Luggenh\"{o}lscher}}]{donko09:131501}
\bibinfo{author}{\bibfnamefont{Z.}~\bibnamefont{Donk\'{o}}},
  \bibinfo{author}{\bibfnamefont{J.}~\bibnamefont{Schulze}},
  \bibinfo{author}{\bibfnamefont{U.}~\bibnamefont{Czarnetzki}},
  \bibnamefont{and}
  \bibinfo{author}{\bibfnamefont{D.}~\bibnamefont{Luggenh\"{o}lscher}},
  \bibinfo{journal}{Applied Physics Letters} \textbf{\bibinfo{volume}{94}},
  \bibinfo{eid}{131501} (\bibinfo{year}{2009}).

\bibitem[{\citenamefont{Sch{\"u}ngel et~al.}(2015)\citenamefont{Sch{\"u}ngel,
  Brandt, Donk{\'o}, Korolov, Derzsi, and Schulze}}]{schungel15:044009}
\bibinfo{author}{\bibfnamefont{E.}~\bibnamefont{Sch{\"u}ngel}},
  \bibinfo{author}{\bibfnamefont{S.}~\bibnamefont{Brandt}},
  \bibinfo{author}{\bibfnamefont{Z.}~\bibnamefont{Donk{\'o}}},
  \bibinfo{author}{\bibfnamefont{I.}~\bibnamefont{Korolov}},
  \bibinfo{author}{\bibfnamefont{A.}~\bibnamefont{Derzsi}}, \bibnamefont{and}
  \bibinfo{author}{\bibfnamefont{J.}~\bibnamefont{Schulze}},
  \bibinfo{journal}{Plasma Sources Science and Technology}
  \textbf{\bibinfo{volume}{24}}, \bibinfo{pages}{044009}
  (\bibinfo{year}{2015}).

\bibitem[{\citenamefont{Wilczek et~al.}(2016)\citenamefont{Wilczek,
  Trieschmann, Eremin, Brinkmann, Schulze, Sch{\"u}ngel, Derzsi, olov,
  Hartmann, Donk{\'o} et~al.}}]{wilczek16:063514}
\bibinfo{author}{\bibfnamefont{S.}~\bibnamefont{Wilczek}},
  \bibinfo{author}{\bibfnamefont{J.}~\bibnamefont{Trieschmann}},
  \bibinfo{author}{\bibfnamefont{D.}~\bibnamefont{Eremin}},
  \bibinfo{author}{\bibfnamefont{R.~P.} \bibnamefont{Brinkmann}},
  \bibinfo{author}{\bibfnamefont{J.}~\bibnamefont{Schulze}},
  \bibinfo{author}{\bibfnamefont{E.}~\bibnamefont{Sch{\"u}ngel}},
  \bibinfo{author}{\bibfnamefont{A.}~\bibnamefont{Derzsi}},
  \bibinfo{author}{\bibfnamefont{I.~K.} \bibnamefont{olov}},
  \bibinfo{author}{\bibfnamefont{P.}~\bibnamefont{Hartmann}},
  \bibinfo{author}{\bibfnamefont{Z.}~\bibnamefont{Donk{\'o}}},
  \bibnamefont{et~al.}, \bibinfo{journal}{Physics of Plasmas}
  \textbf{\bibinfo{volume}{23}}, \bibinfo{pages}{063514}
  (\bibinfo{year}{2016}).

\bibitem[{\citenamefont{Belenguer and Boeuf}(1990)}]{belenguer90:4447}
\bibinfo{author}{\bibfnamefont{P.}~\bibnamefont{Belenguer}} \bibnamefont{and}
  \bibinfo{author}{\bibfnamefont{J.}~\bibnamefont{Boeuf}},
  \bibinfo{journal}{Physical Review A} \textbf{\bibinfo{volume}{41}},
  \bibinfo{pages}{4447} (\bibinfo{year}{1990}).

\bibitem[{\citenamefont{Schulze et~al.}(2011)\citenamefont{Schulze, Derzsi,
  Dittmann, Hemke, Meichsner, and Donk{\'o}}}]{schulze11:275001}
\bibinfo{author}{\bibfnamefont{J.}~\bibnamefont{Schulze}},
  \bibinfo{author}{\bibfnamefont{A.}~\bibnamefont{Derzsi}},
  \bibinfo{author}{\bibfnamefont{K.}~\bibnamefont{Dittmann}},
  \bibinfo{author}{\bibfnamefont{T.}~\bibnamefont{Hemke}},
  \bibinfo{author}{\bibfnamefont{J.}~\bibnamefont{Meichsner}},
  \bibnamefont{and}
  \bibinfo{author}{\bibfnamefont{Z.}~\bibnamefont{Donk{\'o}}},
  \bibinfo{journal}{Physical Review Letters} \textbf{\bibinfo{volume}{107}},
  \bibinfo{pages}{275001} (\bibinfo{year}{2011}).

\bibitem[{\citenamefont{Derzsi et~al.}(2015)\citenamefont{Derzsi, Sch{\"u}ngel,
  Donk{\'o}, and Schulze}}]{derzsi15:346}
\bibinfo{author}{\bibfnamefont{A.}~\bibnamefont{Derzsi}},
  \bibinfo{author}{\bibfnamefont{E.}~\bibnamefont{Sch{\"u}ngel}},
  \bibinfo{author}{\bibfnamefont{Z.}~\bibnamefont{Donk{\'o}}},
  \bibnamefont{and} \bibinfo{author}{\bibfnamefont{J.}~\bibnamefont{Schulze}},
  \bibinfo{journal}{Open Chemistry} \textbf{\bibinfo{volume}{13}},
  \bibinfo{pages}{346} (\bibinfo{year}{2015}).

\bibitem[{\citenamefont{Pulfrey et~al.}(1973)\citenamefont{Pulfrey, Hathorn,
  and Young}}]{pulfrey73:1529}
\bibinfo{author}{\bibfnamefont{D.~L.} \bibnamefont{Pulfrey}},
  \bibinfo{author}{\bibfnamefont{F.~G.~M.} \bibnamefont{Hathorn}},
  \bibnamefont{and} \bibinfo{author}{\bibfnamefont{L.}~\bibnamefont{Young}},
  \bibinfo{journal}{Journal of the Electrochemical Society}
  \textbf{\bibinfo{volume}{120}}, \bibinfo{pages}{1529} (\bibinfo{year}{1973}).

\bibitem[{\citenamefont{Kawai et~al.}(1994)\citenamefont{Kawai, Konishi,
  Watanabe, and Ohmi}}]{kawai94:2223}
\bibinfo{author}{\bibfnamefont{Y.}~\bibnamefont{Kawai}},
  \bibinfo{author}{\bibfnamefont{N.}~\bibnamefont{Konishi}},
  \bibinfo{author}{\bibfnamefont{J.}~\bibnamefont{Watanabe}}, \bibnamefont{and}
  \bibinfo{author}{\bibfnamefont{T.}~\bibnamefont{Ohmi}},
  \bibinfo{journal}{Applied Physics Letters} \textbf{\bibinfo{volume}{64}},
  \bibinfo{pages}{2223} (\bibinfo{year}{1994}).

\bibitem[{\citenamefont{Hess}(1999)}]{hess99:127}
\bibinfo{author}{\bibfnamefont{D.~W.} \bibnamefont{Hess}},
  \bibinfo{journal}{IBM Journal of Research and Development}
  \textbf{\bibinfo{volume}{43}}, \bibinfo{pages}{127} (\bibinfo{year}{1999}).

\bibitem[{\citenamefont{Tolliver}(1984)}]{tolliver84:1}
\bibinfo{author}{\bibfnamefont{D.~L.} \bibnamefont{Tolliver}}, in
  \emph{\bibinfo{booktitle}{VLSI Electronics: Microstructure Science, vol. 8}},
  edited by \bibinfo{editor}{\bibfnamefont{N.~G.} \bibnamefont{Einspruch}}
  \bibnamefont{and} \bibinfo{editor}{\bibfnamefont{D.~M.} \bibnamefont{Brown}}
  (\bibinfo{publisher}{Academic Press}, \bibinfo{address}{Orlando},
  \bibinfo{year}{1984}), pp. \bibinfo{pages}{1--24}.

\bibitem[{\citenamefont{Hartney et~al.}(1989)\citenamefont{Hartney, Hess, and
  Soane}}]{hartney89:1}
\bibinfo{author}{\bibfnamefont{M.~A.} \bibnamefont{Hartney}},
  \bibinfo{author}{\bibfnamefont{D.~W.} \bibnamefont{Hess}}, \bibnamefont{and}
  \bibinfo{author}{\bibfnamefont{D.~S.} \bibnamefont{Soane}},
  \bibinfo{journal}{Journal of Vacuum Science and Technology B}
  \textbf{\bibinfo{volume}{7}}, \bibinfo{pages}{1} (\bibinfo{year}{1989}).

\bibitem[{\citenamefont{Vesel and Mozetic}(2012)}]{vesel12:634}
\bibinfo{author}{\bibfnamefont{A.}~\bibnamefont{Vesel}} \bibnamefont{and}
  \bibinfo{author}{\bibfnamefont{M.}~\bibnamefont{Mozetic}},
  \bibinfo{journal}{Vacuum} \textbf{\bibinfo{volume}{86}}, \bibinfo{pages}{634}
  (\bibinfo{year}{2012}).

\bibitem[{\citenamefont{Chashmejahanbin
  et~al.}(2014)\citenamefont{Chashmejahanbin, Salimi, and {Ershad
  Langroudi}}}]{chashmejahanbin14:44}
\bibinfo{author}{\bibfnamefont{M.~R.} \bibnamefont{Chashmejahanbin}},
  \bibinfo{author}{\bibfnamefont{A.}~\bibnamefont{Salimi}}, \bibnamefont{and}
  \bibinfo{author}{\bibfnamefont{A.}~\bibnamefont{{Ershad Langroudi}}},
  \bibinfo{journal}{International Journal of Adhesion and Adhesives}
  \textbf{\bibinfo{volume}{49}}, \bibinfo{pages}{44} (\bibinfo{year}{2014}).

\bibitem[{\citenamefont{Vesel and Mozetic}(2017)}]{vesel17:293001}
\bibinfo{author}{\bibfnamefont{A.}~\bibnamefont{Vesel}} \bibnamefont{and}
  \bibinfo{author}{\bibfnamefont{M.}~\bibnamefont{Mozetic}},
  \bibinfo{journal}{Journal of Physics D: Applied Physics}
  \textbf{\bibinfo{volume}{50}}, \bibinfo{pages}{293001}
  (\bibinfo{year}{2017}).

\bibitem[{\citenamefont{Gudmundsson et~al.}(2001)\citenamefont{Gudmundsson,
  Kouznetsov, Patel, and Lieberman}}]{gudmundsson01:1100}
\bibinfo{author}{\bibfnamefont{J.~T.} \bibnamefont{Gudmundsson}},
  \bibinfo{author}{\bibfnamefont{I.~G.} \bibnamefont{Kouznetsov}},
  \bibinfo{author}{\bibfnamefont{K.~K.} \bibnamefont{Patel}}, \bibnamefont{and}
  \bibinfo{author}{\bibfnamefont{M.~A.} \bibnamefont{Lieberman}},
  \bibinfo{journal}{Journal of Physics D: Applied Physics}
  \textbf{\bibinfo{volume}{34}}, \bibinfo{pages}{1100} (\bibinfo{year}{2001}).

\bibitem[{\citenamefont{Gudmundsson and
  Thorsteinsson}(2007)}]{gudmundsson07:399}
\bibinfo{author}{\bibfnamefont{J.~T.} \bibnamefont{Gudmundsson}}
  \bibnamefont{and} \bibinfo{author}{\bibfnamefont{E.~G.}
  \bibnamefont{Thorsteinsson}}, \bibinfo{journal}{Plasma Sources Science and
  Technology} \textbf{\bibinfo{volume}{16}}, \bibinfo{pages}{399}
  (\bibinfo{year}{2007}).

\bibitem[{\citenamefont{Toneli et~al.}(2015)\citenamefont{Toneli, Pessoa,
  Roberto, and Gudmundsson}}]{toneli15:325202}
\bibinfo{author}{\bibfnamefont{D.~A.} \bibnamefont{Toneli}},
  \bibinfo{author}{\bibfnamefont{R.~S.} \bibnamefont{Pessoa}},
  \bibinfo{author}{\bibfnamefont{M.}~\bibnamefont{Roberto}}, \bibnamefont{and}
  \bibinfo{author}{\bibfnamefont{J.~T.} \bibnamefont{Gudmundsson}},
  \bibinfo{journal}{Journal of Physics D: Applied Physics}
  \textbf{\bibinfo{volume}{48}}, \bibinfo{pages}{325202}
  (\bibinfo{year}{2015}).

\bibitem[{\citenamefont{Gudmundsson and
  Lieberman}(2015)}]{gudmundsson15:035016}
\bibinfo{author}{\bibfnamefont{J.~T.} \bibnamefont{Gudmundsson}}
  \bibnamefont{and} \bibinfo{author}{\bibfnamefont{M.~A.}
  \bibnamefont{Lieberman}}, \bibinfo{journal}{Plasma Sources Science and
  Technology} \textbf{\bibinfo{volume}{24}}, \bibinfo{pages}{035016}
  (\bibinfo{year}{2015}).

\bibitem[{\citenamefont{Gudmundsson and
  Vent{\'e}jou}(2015)}]{gudmundsson15:153302}
\bibinfo{author}{\bibfnamefont{J.~T.} \bibnamefont{Gudmundsson}}
  \bibnamefont{and}
  \bibinfo{author}{\bibfnamefont{B.}~\bibnamefont{Vent{\'e}jou}},
  \bibinfo{journal}{Journal of Applied Physics} \textbf{\bibinfo{volume}{118}},
  \bibinfo{pages}{153302} (\bibinfo{year}{2015}).

\bibitem[{\citenamefont{Hannesdottir and
  Gudmundsson}(2016)}]{hannesdottir16:055002}
\bibinfo{author}{\bibfnamefont{H.}~\bibnamefont{Hannesdottir}}
  \bibnamefont{and} \bibinfo{author}{\bibfnamefont{J.~T.}
  \bibnamefont{Gudmundsson}}, \bibinfo{journal}{Plasma Sources Science and
  Technology} \textbf{\bibinfo{volume}{25}}, \bibinfo{pages}{055002}
  (\bibinfo{year}{2016}).

\bibitem[{\citenamefont{Gudmundsson and
  Hannesdottir}(2017)}]{gudmundsson17:120001}
\bibinfo{author}{\bibfnamefont{J.~T.} \bibnamefont{Gudmundsson}}
  \bibnamefont{and}
  \bibinfo{author}{\bibfnamefont{H.}~\bibnamefont{Hannesdottir}},
  \bibinfo{journal}{AIP Conference Proceedings}
  \textbf{\bibinfo{volume}{1811}}, \bibinfo{pages}{120001}
  (\bibinfo{year}{2017}).

\bibitem[{\citenamefont{Hannesdottir and
  Gudmundsson}(2017)}]{hannesdottir17:175201}
\bibinfo{author}{\bibfnamefont{H.}~\bibnamefont{Hannesdottir}}
  \bibnamefont{and} \bibinfo{author}{\bibfnamefont{J.~T.}
  \bibnamefont{Gudmundsson}}, \bibinfo{journal}{Journal of Physics D: Applied
  Physics} \textbf{\bibinfo{volume}{50}}, \bibinfo{pages}{175201}
  (\bibinfo{year}{2017}).

\bibitem[{\citenamefont{Gudmundsson and
  Snorrason}(2017)}]{gudmundsson17:193302}
\bibinfo{author}{\bibfnamefont{J.~T.} \bibnamefont{Gudmundsson}}
  \bibnamefont{and} \bibinfo{author}{\bibfnamefont{D.~I.}
  \bibnamefont{Snorrason}}, \bibinfo{journal}{Journal of Applied Physics}
  \textbf{\bibinfo{volume}{122}}, \bibinfo{pages}{193302}
  (\bibinfo{year}{2017}).

\bibitem[{\citenamefont{Gudmundsson et~al.}(2018)\citenamefont{Gudmundsson,
  Snorrason, and Hannesdottir}}]{gudmundsson18:025009}
\bibinfo{author}{\bibfnamefont{J.~T.} \bibnamefont{Gudmundsson}},
  \bibinfo{author}{\bibfnamefont{D.~I.} \bibnamefont{Snorrason}},
  \bibnamefont{and}
  \bibinfo{author}{\bibfnamefont{H.}~\bibnamefont{Hannesdottir}},
  \bibinfo{journal}{Plasma Sources Science and Technology}
  \textbf{\bibinfo{volume}{27}}, \bibinfo{pages}{025009}
  (\bibinfo{year}{2018}).

\bibitem[{\citenamefont{Greb et~al.}(2015)\citenamefont{Greb, Gibson, Niemi,
  {O'C}onnell, and Gans}}]{greb15:044003}
\bibinfo{author}{\bibfnamefont{A.}~\bibnamefont{Greb}},
  \bibinfo{author}{\bibfnamefont{A.~R.} \bibnamefont{Gibson}},
  \bibinfo{author}{\bibfnamefont{K.}~\bibnamefont{Niemi}},
  \bibinfo{author}{\bibfnamefont{D.}~\bibnamefont{{O'C}onnell}},
  \bibnamefont{and} \bibinfo{author}{\bibfnamefont{T.}~\bibnamefont{Gans}},
  \bibinfo{journal}{Plasma Sources Science and Technology}
  \textbf{\bibinfo{volume}{24}}, \bibinfo{pages}{044003}
  (\bibinfo{year}{2015}).

\bibitem[{\citenamefont{Derzsi et~al.}(2016)\citenamefont{Derzsi, Lafleur,
  Booth, Korolov, and Donk{\'o}}}]{derzsi16:015004}
\bibinfo{author}{\bibfnamefont{A.}~\bibnamefont{Derzsi}},
  \bibinfo{author}{\bibfnamefont{T.}~\bibnamefont{Lafleur}},
  \bibinfo{author}{\bibfnamefont{J.-P.} \bibnamefont{Booth}},
  \bibinfo{author}{\bibfnamefont{I.}~\bibnamefont{Korolov}}, \bibnamefont{and}
  \bibinfo{author}{\bibfnamefont{Z.}~\bibnamefont{Donk{\'o}}},
  \bibinfo{journal}{Plasma Sources Science and Technology}
  \textbf{\bibinfo{volume}{25}}, \bibinfo{pages}{015004}
  (\bibinfo{year}{2016}).

\bibitem[{\citenamefont{Derzsi et~al.}(2017)\citenamefont{Derzsi, Bruneau,
  Gibson, Johnson, {O'C}onnell, Gans, Booth, and Donk{\'o}}}]{derzsi17:034002}
\bibinfo{author}{\bibfnamefont{A.}~\bibnamefont{Derzsi}},
  \bibinfo{author}{\bibfnamefont{B.}~\bibnamefont{Bruneau}},
  \bibinfo{author}{\bibfnamefont{A.}~\bibnamefont{Gibson}},
  \bibinfo{author}{\bibfnamefont{E.}~\bibnamefont{Johnson}},
  \bibinfo{author}{\bibfnamefont{D.}~\bibnamefont{{O'C}onnell}},
  \bibinfo{author}{\bibfnamefont{T.}~\bibnamefont{Gans}},
  \bibinfo{author}{\bibfnamefont{J.-P.} \bibnamefont{Booth}}, \bibnamefont{and}
  \bibinfo{author}{\bibfnamefont{Z.}~\bibnamefont{Donk{\'o}}},
  \bibinfo{journal}{Plasma Sources Science and Technology}
  \textbf{\bibinfo{volume}{26}}, \bibinfo{pages}{034002}
  (\bibinfo{year}{2017}).

\bibitem[{\citenamefont{Gibson and Gans}(2017)}]{gibson17:115007}
\bibinfo{author}{\bibfnamefont{A.~R.} \bibnamefont{Gibson}} \bibnamefont{and}
  \bibinfo{author}{\bibfnamefont{T.}~\bibnamefont{Gans}},
  \bibinfo{journal}{Plasma Sources Science and Technology}
  \textbf{\bibinfo{volume}{26}}, \bibinfo{pages}{115007}
  (\bibinfo{year}{2017}).

\bibitem[{\citenamefont{Hammel and Verboncoeur}(2003)}]{hammel03:66}
\bibinfo{author}{\bibfnamefont{J.}~\bibnamefont{Hammel}} \bibnamefont{and}
  \bibinfo{author}{\bibfnamefont{J.~P.} \bibnamefont{Verboncoeur}},
  \bibinfo{journal}{Bulletin of the American Physical Society}
  \textbf{\bibinfo{volume}{48}}, \bibinfo{pages}{66} (\bibinfo{year}{2003}).

\bibitem[{\citenamefont{Verboncoeur et~al.}(1995)\citenamefont{Verboncoeur,
  Langdon, and Gladd}}]{verboncoeur95:199}
\bibinfo{author}{\bibfnamefont{J.~P.} \bibnamefont{Verboncoeur}},
  \bibinfo{author}{\bibfnamefont{A.~B.} \bibnamefont{Langdon}},
  \bibnamefont{and} \bibinfo{author}{\bibfnamefont{N.~T.} \bibnamefont{Gladd}},
  \bibinfo{journal}{Computer Physics Communications}
  \textbf{\bibinfo{volume}{87}}, \bibinfo{pages}{199} (\bibinfo{year}{1995}).

\bibitem[{\citenamefont{Gudmundsson et~al.}(2013)\citenamefont{Gudmundsson,
  Kawamura, and Lieberman}}]{gudmundsson13:035011}
\bibinfo{author}{\bibfnamefont{J.~T.} \bibnamefont{Gudmundsson}},
  \bibinfo{author}{\bibfnamefont{E.}~\bibnamefont{Kawamura}}, \bibnamefont{and}
  \bibinfo{author}{\bibfnamefont{M.~A.} \bibnamefont{Lieberman}},
  \bibinfo{journal}{Plasma Sources Science and Technology}
  \textbf{\bibinfo{volume}{22}}, \bibinfo{pages}{035011}
  (\bibinfo{year}{2013}).

\bibitem[{\citenamefont{Midey et~al.}(2008)\citenamefont{Midey, Dotan, and
  Viggiano}}]{midey08:3040}
\bibinfo{author}{\bibfnamefont{A.}~\bibnamefont{Midey}},
  \bibinfo{author}{\bibfnamefont{I.}~\bibnamefont{Dotan}}, \bibnamefont{and}
  \bibinfo{author}{\bibfnamefont{A.~A.} \bibnamefont{Viggiano}},
  \bibinfo{journal}{Journal of Physical Chemistry A}
  \textbf{\bibinfo{volume}{113}}, \bibinfo{pages}{3040} (\bibinfo{year}{2008}).

\bibitem[{\citenamefont{Lichtenberg et~al.}(1994)\citenamefont{Lichtenberg,
  Vahedi, Lieberman, and Rognlien}}]{lichtenberg94:2339}
\bibinfo{author}{\bibfnamefont{A.~J.} \bibnamefont{Lichtenberg}},
  \bibinfo{author}{\bibfnamefont{V.}~\bibnamefont{Vahedi}},
  \bibinfo{author}{\bibfnamefont{M.~A.} \bibnamefont{Lieberman}},
  \bibnamefont{and} \bibinfo{author}{\bibfnamefont{T.}~\bibnamefont{Rognlien}},
  \bibinfo{journal}{Journal of Applied Physics} \textbf{\bibinfo{volume}{75}},
  \bibinfo{pages}{2339 } (\bibinfo{year}{1994}).

\bibitem[{\citenamefont{Birdsall}(1991)}]{birdsall91:65}
\bibinfo{author}{\bibfnamefont{C.}~\bibnamefont{Birdsall}},
  \bibinfo{journal}{IEEE Transactions on Plasma Science}
  \textbf{\bibinfo{volume}{19}}, \bibinfo{pages}{65} (\bibinfo{year}{1991}).

\bibitem[{\citenamefont{Kawamura et~al.}(2000)\citenamefont{Kawamura, Birdsall,
  and Vahedi}}]{kawamura00:413}
\bibinfo{author}{\bibfnamefont{E.}~\bibnamefont{Kawamura}},
  \bibinfo{author}{\bibfnamefont{C.~K.} \bibnamefont{Birdsall}},
  \bibnamefont{and} \bibinfo{author}{\bibfnamefont{V.}~\bibnamefont{Vahedi}},
  \bibinfo{journal}{Plasma Sources Science and Technology}
  \textbf{\bibinfo{volume}{9}}, \bibinfo{pages}{413} (\bibinfo{year}{2000}).

\bibitem[{\citenamefont{Nguyen}(2006)}]{nguyen06}
\bibinfo{author}{\bibfnamefont{C.}~\bibnamefont{Nguyen}}, Master's thesis,
  \bibinfo{school}{University of California at Berkeley}
  (\bibinfo{year}{2006}).

\bibitem[{\citenamefont{Lim}(2007)}]{lim07}
\bibinfo{author}{\bibfnamefont{C.-H.} \bibnamefont{Lim}}, Ph.D. thesis,
  \bibinfo{school}{University of California at Berkeley}
  (\bibinfo{year}{2007}).

\bibitem[{\citenamefont{Miller and Combi}(1994)}]{miller94:1735}
\bibinfo{author}{\bibfnamefont{R.~H.} \bibnamefont{Miller}} \bibnamefont{and}
  \bibinfo{author}{\bibfnamefont{M.~R.} \bibnamefont{Combi}},
  \bibinfo{journal}{Geophysical Research Letters}
  \textbf{\bibinfo{volume}{21}}, \bibinfo{pages}{1735} (\bibinfo{year}{1994}).

\bibitem[{\citenamefont{Booth and Sadeghi}(1991)}]{booth91:611}
\bibinfo{author}{\bibfnamefont{J.~P.} \bibnamefont{Booth}} \bibnamefont{and}
  \bibinfo{author}{\bibfnamefont{N.}~\bibnamefont{Sadeghi}},
  \bibinfo{journal}{Journal of Applied Physics} \textbf{\bibinfo{volume}{70}},
  \bibinfo{pages}{611} (\bibinfo{year}{1991}).

\bibitem[{\citenamefont{Du et~al.}(2011)\citenamefont{Du, Leng, Yang, Sha, and
  Zhang}}]{du11:256}
\bibinfo{author}{\bibfnamefont{S.}~\bibnamefont{Du}},
  \bibinfo{author}{\bibfnamefont{J.}~\bibnamefont{Leng}},
  \bibinfo{author}{\bibfnamefont{H.}~\bibnamefont{Yang}},
  \bibinfo{author}{\bibfnamefont{G.}~\bibnamefont{Sha}}, \bibnamefont{and}
  \bibinfo{author}{\bibfnamefont{C.}~\bibnamefont{Zhang}},
  \bibinfo{journal}{Chinese Journal of Chemical Physics}
  \textbf{\bibinfo{volume}{24}}, \bibinfo{pages}{256} (\bibinfo{year}{2011}).

\bibitem[{\citenamefont{Sharpless and Slanger}(1989)}]{sharpless89:7947}
\bibinfo{author}{\bibfnamefont{R.~L.} \bibnamefont{Sharpless}}
  \bibnamefont{and} \bibinfo{author}{\bibfnamefont{T.~G.}
  \bibnamefont{Slanger}}, \bibinfo{journal}{Journal of Chemical Physics}
  \textbf{\bibinfo{volume}{91}}, \bibinfo{pages}{7947 } (\bibinfo{year}{1989}).

\bibitem[{\citenamefont{O'{B}rien and Myers}(1970)}]{obrien70:3832}
\bibinfo{author}{\bibfnamefont{R.~J.} \bibnamefont{O'{B}rien}}
  \bibnamefont{and} \bibinfo{author}{\bibfnamefont{G.~H.} \bibnamefont{Myers}},
  \bibinfo{journal}{Journal of Chemical Physics} \textbf{\bibinfo{volume}{53}},
  \bibinfo{pages}{3832} (\bibinfo{year}{1970}).

\bibitem[{\citenamefont{Gordiets et~al.}(1995)\citenamefont{Gordiets, Ferreira,
  Guerra, Loureiro, Nahorny, Pagnon, Touzeau, and Vialle}}]{gordiets95:750}
\bibinfo{author}{\bibfnamefont{B.}~\bibnamefont{Gordiets}},
  \bibinfo{author}{\bibfnamefont{C.}~\bibnamefont{Ferreira}},
  \bibinfo{author}{\bibfnamefont{V.}~\bibnamefont{Guerra}},
  \bibinfo{author}{\bibfnamefont{J.}~\bibnamefont{Loureiro}},
  \bibinfo{author}{\bibfnamefont{J.}~\bibnamefont{Nahorny}},
  \bibinfo{author}{\bibfnamefont{D.}~\bibnamefont{Pagnon}},
  \bibinfo{author}{\bibfnamefont{M.}~\bibnamefont{Touzeau}}, \bibnamefont{and}
  \bibinfo{author}{\bibfnamefont{M.}~\bibnamefont{Vialle}},
  \bibinfo{journal}{IEEE Transactions on Plasma Science}
  \textbf{\bibinfo{volume}{23}}, \bibinfo{pages}{750} (\bibinfo{year}{1995}).

\bibitem[{\citenamefont{Kutasi et~al.}(2010)\citenamefont{Kutasi, Guerra, and
  S{\'a}}}]{kutasi10:175201}
\bibinfo{author}{\bibfnamefont{K.}~\bibnamefont{Kutasi}},
  \bibinfo{author}{\bibfnamefont{V.}~\bibnamefont{Guerra}}, \bibnamefont{and}
  \bibinfo{author}{\bibfnamefont{P.}~\bibnamefont{S{\'a}}},
  \bibinfo{journal}{Journal of Physics D: Applied Physics}
  \textbf{\bibinfo{volume}{43}}, \bibinfo{pages}{175201}
  (\bibinfo{year}{2010}).

\bibitem[{\citenamefont{Greb et~al.}(2013)\citenamefont{Greb, Niemi,
  {O'Connell}, and Gans}}]{greb13:244101}
\bibinfo{author}{\bibfnamefont{A.}~\bibnamefont{Greb}},
  \bibinfo{author}{\bibfnamefont{K.}~\bibnamefont{Niemi}},
  \bibinfo{author}{\bibfnamefont{D.}~\bibnamefont{{O'Connell}}},
  \bibnamefont{and} \bibinfo{author}{\bibfnamefont{T.}~\bibnamefont{Gans}},
  \bibinfo{journal}{Applied Physics Letters} \textbf{\bibinfo{volume}{103}},
  \bibinfo{pages}{244101} (\bibinfo{year}{2013}).

\bibitem[{\citenamefont{Crannage et~al.}(1993)\citenamefont{Crannage, Dorko,
  Johnson, and Whitefield}}]{crannage93:267}
\bibinfo{author}{\bibfnamefont{R.~P.} \bibnamefont{Crannage}},
  \bibinfo{author}{\bibfnamefont{E.~A.} \bibnamefont{Dorko}},
  \bibinfo{author}{\bibfnamefont{D.~E.} \bibnamefont{Johnson}},
  \bibnamefont{and} \bibinfo{author}{\bibfnamefont{P.~D.}
  \bibnamefont{Whitefield}}, \bibinfo{journal}{Chemical Physics}
  \textbf{\bibinfo{volume}{169}}, \bibinfo{pages}{267 } (\bibinfo{year}{1993}).

\bibitem[{\citenamefont{Perram et~al.}(1992)\citenamefont{Perram, Determan,
  Dorian, Lowe, and Thompson}}]{perram92:427}
\bibinfo{author}{\bibfnamefont{G.}~\bibnamefont{Perram}},
  \bibinfo{author}{\bibfnamefont{D.}~\bibnamefont{Determan}},
  \bibinfo{author}{\bibfnamefont{J.}~\bibnamefont{Dorian}},
  \bibinfo{author}{\bibfnamefont{B.}~\bibnamefont{Lowe}}, \bibnamefont{and}
  \bibinfo{author}{\bibfnamefont{T.~L.} \bibnamefont{Thompson}},
  \bibinfo{journal}{Chemical Physics} \textbf{\bibinfo{volume}{162}},
  \bibinfo{pages}{427} (\bibinfo{year}{1992}).

\bibitem[{\citenamefont{Steer et~al.}(1969)\citenamefont{Steer, Ackerman, and
  Pitts}}]{steer69:843}
\bibinfo{author}{\bibfnamefont{R.~P.} \bibnamefont{Steer}},
  \bibinfo{author}{\bibfnamefont{R.~A.} \bibnamefont{Ackerman}},
  \bibnamefont{and} \bibinfo{author}{\bibfnamefont{J.~N.} \bibnamefont{Pitts}},
  \bibinfo{journal}{Journal of Chemical Physics} \textbf{\bibinfo{volume}{51}},
  \bibinfo{pages}{843} (\bibinfo{year}{1969}).

\bibitem[{\citenamefont{Arnold and Ogryzlo}(1967)}]{arnold67:2053}
\bibinfo{author}{\bibfnamefont{S.~J.} \bibnamefont{Arnold}} \bibnamefont{and}
  \bibinfo{author}{\bibfnamefont{E.~A.} \bibnamefont{Ogryzlo}},
  \bibinfo{journal}{Canadian Journal of Physics} \textbf{\bibinfo{volume}{45}},
  \bibinfo{pages}{2053} (\bibinfo{year}{1967}).

\bibitem[{\citenamefont{Clark and Wayne}(1969)}]{clark69:93}
\bibinfo{author}{\bibfnamefont{I.~D.} \bibnamefont{Clark}} \bibnamefont{and}
  \bibinfo{author}{\bibfnamefont{R.~P.} \bibnamefont{Wayne}},
  \bibinfo{journal}{Chemical Physics Letters} \textbf{\bibinfo{volume}{3}},
  \bibinfo{pages}{93 } (\bibinfo{year}{1969}).

\bibitem[{\citenamefont{Izod and Wayne}(1968)}]{izod68:81}
\bibinfo{author}{\bibfnamefont{T.~P.~J.} \bibnamefont{Izod}} \bibnamefont{and}
  \bibinfo{author}{\bibfnamefont{R.~P.} \bibnamefont{Wayne}},
  \bibinfo{journal}{Proceedings of the Royal Society of London. Series A,
  Mathematical and Physical Sciences} \textbf{\bibinfo{volume}{308}},
  \bibinfo{pages}{81} (\bibinfo{year}{1968}).

\bibitem[{\citenamefont{Leiss et~al.}(1978)\citenamefont{Leiss, Schurath,
  Becker, and Fink}}]{leiss78:211}
\bibinfo{author}{\bibfnamefont{A.}~\bibnamefont{Leiss}},
  \bibinfo{author}{\bibfnamefont{U.}~\bibnamefont{Schurath}},
  \bibinfo{author}{\bibfnamefont{K.~H.} \bibnamefont{Becker}},
  \bibnamefont{and} \bibinfo{author}{\bibfnamefont{E.~H.} \bibnamefont{Fink}},
  \bibinfo{journal}{Journal of Photochemistry} \textbf{\bibinfo{volume}{8}},
  \bibinfo{pages}{211} (\bibinfo{year}{1978}).

\bibitem[{\citenamefont{Ryskin and Shub}(1981)}]{ryskin81:41}
\bibinfo{author}{\bibfnamefont{M.~E.} \bibnamefont{Ryskin}} \bibnamefont{and}
  \bibinfo{author}{\bibfnamefont{B.~R.} \bibnamefont{Shub}},
  \bibinfo{journal}{Reaction Kinetics and Catalysis Letters}
  \textbf{\bibinfo{volume}{17}}, \bibinfo{pages}{41} (\bibinfo{year}{1981}).

\bibitem[{\citenamefont{Thorsteinsson and
  Gudmundsson}(2010)}]{thorsteinsson10:055008}
\bibinfo{author}{\bibfnamefont{E.~G.} \bibnamefont{Thorsteinsson}}
  \bibnamefont{and} \bibinfo{author}{\bibfnamefont{J.~T.}
  \bibnamefont{Gudmundsson}}, \bibinfo{journal}{Plasma Sources Science and
  Technology} \textbf{\bibinfo{volume}{19}}, \bibinfo{pages}{055008}
  (\bibinfo{year}{2010}).

\bibitem[{\citenamefont{Jonsson}(2018)}]{jonsson18bt}
\bibinfo{author}{\bibfnamefont{R.~D.~B.} \bibnamefont{Jonsson}},
  \emph{\bibinfo{title}{{\rm B.S.~project, University of Iceland, Reykjavik,
  Iceland, {\tt http://hdl.handle.net/1946/29537}}}} (\bibinfo{year}{2018}).

\bibitem[{\citenamefont{{O'C}onnell et~al.}(2007)\citenamefont{{O'C}onnell,
  Gans, Vender, Czarnetzki, and Boswell}}]{oconnell07:034505}
\bibinfo{author}{\bibfnamefont{D.}~\bibnamefont{{O'C}onnell}},
  \bibinfo{author}{\bibfnamefont{T.}~\bibnamefont{Gans}},
  \bibinfo{author}{\bibfnamefont{D.}~\bibnamefont{Vender}},
  \bibinfo{author}{\bibfnamefont{U.}~\bibnamefont{Czarnetzki}},
  \bibnamefont{and} \bibinfo{author}{\bibfnamefont{R.}~\bibnamefont{Boswell}},
  \bibinfo{journal}{Physics of Plasmas} \textbf{\bibinfo{volume}{14}},
  \bibinfo{eid}{034505} (\bibinfo{year}{2007}).

\bibitem[{\citenamefont{Berezhnoj et~al.}(2000)\citenamefont{Berezhnoj, Shin,
  Buddemeier, and Kaganovich}}]{berezhnoj00:800}
\bibinfo{author}{\bibfnamefont{S.~V.} \bibnamefont{Berezhnoj}},
  \bibinfo{author}{\bibfnamefont{C.~B.} \bibnamefont{Shin}},
  \bibinfo{author}{\bibfnamefont{U.}~\bibnamefont{Buddemeier}},
  \bibnamefont{and}
  \bibinfo{author}{\bibfnamefont{I.}~\bibnamefont{Kaganovich}},
  \bibinfo{journal}{Applied Physics Letters} \textbf{\bibinfo{volume}{77}},
  \bibinfo{pages}{800} (\bibinfo{year}{2000}).

\bibitem[{\citenamefont{Katsch et~al.}(2000)\citenamefont{Katsch, Sturm,
  Quandt, and D{\"o}bele}}]{katsch00:323}
\bibinfo{author}{\bibfnamefont{H.~M.} \bibnamefont{Katsch}},
  \bibinfo{author}{\bibfnamefont{T.}~\bibnamefont{Sturm}},
  \bibinfo{author}{\bibfnamefont{E.}~\bibnamefont{Quandt}}, \bibnamefont{and}
  \bibinfo{author}{\bibfnamefont{H.~F.} \bibnamefont{D{\"o}bele}},
  \bibinfo{journal}{Plasma Sources Science and Technology}
  \textbf{\bibinfo{volume}{9}}, \bibinfo{pages}{323} (\bibinfo{year}{2000}).

\bibitem[{\citenamefont{K{\"u}llig et~al.}(2010)\citenamefont{K{\"u}llig,
  Dittmann, and Meichsner}}]{kullig10:065011}
\bibinfo{author}{\bibfnamefont{C.}~\bibnamefont{K{\"u}llig}},
  \bibinfo{author}{\bibfnamefont{K.}~\bibnamefont{Dittmann}}, \bibnamefont{and}
  \bibinfo{author}{\bibfnamefont{J.}~\bibnamefont{Meichsner}},
  \bibinfo{journal}{Plasma Sources Science and Technology}
  \textbf{\bibinfo{volume}{19}}, \bibinfo{pages}{065011}
  (\bibinfo{year}{2010}).

\bibitem[{\citenamefont{Kaga et~al.}(2001)\citenamefont{Kaga, Kimura, and
  Ohe}}]{kaga01:330}
\bibinfo{author}{\bibfnamefont{K.}~\bibnamefont{Kaga}},
  \bibinfo{author}{\bibfnamefont{T.}~\bibnamefont{Kimura}}, \bibnamefont{and}
  \bibinfo{author}{\bibfnamefont{K.}~\bibnamefont{Ohe}},
  \bibinfo{journal}{Japanese Journal of Applied Physics}
  \textbf{\bibinfo{volume}{40}}, \bibinfo{pages}{330} (\bibinfo{year}{2001}).

\bibitem[{\citenamefont{Kechkar et~al.}(2017)\citenamefont{Kechkar, Swift,
  Kelly, Kumar, Daniels, and Turner}}]{kechkar17:065009}
\bibinfo{author}{\bibfnamefont{S.}~\bibnamefont{Kechkar}},
  \bibinfo{author}{\bibfnamefont{P.}~\bibnamefont{Swift}},
  \bibinfo{author}{\bibfnamefont{S.}~\bibnamefont{Kelly}},
  \bibinfo{author}{\bibfnamefont{S.}~\bibnamefont{Kumar}},
  \bibinfo{author}{\bibfnamefont{S.}~\bibnamefont{Daniels}}, \bibnamefont{and}
  \bibinfo{author}{\bibfnamefont{M.}~\bibnamefont{Turner}},
  \bibinfo{journal}{Plasma Sources Science and Technology}
  \textbf{\bibinfo{volume}{25}}, \bibinfo{pages}{065009}
  (\bibinfo{year}{2017}).

\bibitem[{\citenamefont{Liu et~al.}(2015)\citenamefont{Liu, Liu, Wen, and
  Wang}}]{liu15:034006}
\bibinfo{author}{\bibfnamefont{G.-H.} \bibnamefont{Liu}},
  \bibinfo{author}{\bibfnamefont{Y.-X.} \bibnamefont{Liu}},
  \bibinfo{author}{\bibfnamefont{D.-Q.} \bibnamefont{Wen}}, \bibnamefont{and}
  \bibinfo{author}{\bibfnamefont{Y.-N.} \bibnamefont{Wang}},
  \bibinfo{journal}{Plasma Sources Science and Technology}
  \textbf{\bibinfo{volume}{24}}, \bibinfo{pages}{034006}
  (\bibinfo{year}{2015}).

\end{thebibliography}

%
%

%
%

\end{document}